\documentclass[sn-mathphys,Numbered]{sn-jnl}

\usepackage{graphicx}%
\usepackage{multirow}%
\usepackage{amsmath,amssymb,amsfonts}%
\usepackage{amsthm}%
\usepackage{mathrsfs}%
\usepackage[title]{appendix}%
\usepackage{xcolor}%
\usepackage{textcomp}%
\usepackage{manyfoot}%
\usepackage{booktabs}%
\usepackage{algorithm}%
\usepackage{algorithmicx}%
\usepackage{algpseudocode}%
\usepackage{listings}%
\usepackage{qcircuit}
\usepackage{caption}
\usepackage{subcaption}
\usepackage{hyperref}
\usepackage{physics}
\theoremstyle{thmstyleone}%
\theoremstyle{thmstyletwo}%

\theoremstyle{thmstylethree}%
\begin{document}

\title[Article Title]{VQC-Based Reinforcement Learning with Data Re-uploading: Performance and Trainability}

\author*[1,2]{\fnm{Rodrigo} \sur{Coelho}}\email{rodrigocoelho140@gmail.com}

\author[1,2,3]{\fnm{André} \sur{Sequeira}}\email{andresequeira410@gmail.com}
\equalcont{These authors contributed equally to this work.}

\author[1,2,3]{\fnm{Luís} \sur{Paulo Santos}}\email{psantos@di.uminho.pt}
\equalcont{These authors contributed equally to this work.}

\affil*[1]{\orgdiv{Department of Informatics}, \orgname{University of Minho}, \orgaddress{ \city{Braga}, \country{Portugal}}}

\affil[2]{\orgdiv{HASLab}, \orgname{INESC TEC}, \orgaddress{\city{Braga}, \country{Portugal}}}

\affil[3]{\orgdiv{International Nanotechnology Laboratory (INL)}, \orgaddress{\city{Braga}, \country{Portugal}}}


\abstract{Reinforcement Learning (RL) consists of designing agents that make intelligent decisions without human supervision. When used alongside function approximators such as Neural Networks (NNs), RL is capable of solving extremely complex problems. Deep Q-Learning, a RL algorithm that uses Deep NNs, has been shown to achieve super-human performance in game-related tasks. Nonetheless, it is also possible to use Variational Quantum Circuits (VQCs) as function approximators in RL algorithms. This work empirically studies the performance and trainability of such VQC-based Deep Q-Learning models in classic control benchmark environments. More specifically, we research how data re-uploading affects both these metrics. We show that the magnitude and the variance of the model's gradients remain substantial throughout training even as the number of qubits increases. In fact, both increase considerably in the training's early stages, when the agent needs to learn the most. They decrease later in the training, when the agent should have done most of the learning and started converging to a policy. Thus, even if the probability of being initialized in a Barren Plateau increases exponentially with system size for Hardware-Efficient ansatzes, these results indicate that the VQC-based Deep Q-Learning models may still be able to find large gradients throughout training, allowing for learning.}





\maketitle

\section*{Statements and Declarations}
\subsection*{Conflict of interests}
The authors declare no competing interests.

\pagebreak

\section{Introduction}\label{sec1}
\subsection{Context}

With quantum computers still in the \emph{Noisy Intermediate Scale Quantum (NISQ)} era, algorithms with theoretical complexity advantages over the best-known classical algorithms, such as Shor's algorithm to factor large integers \cite{shor1999polynomial}, can not be implemented to solve practical problems \cite{preskill2018quantum}. Thus, much of nowadays quantum computing research focuses on finding near-term quantum advantages. A field of quantum computing that has attracted considerable interest as one of the most promising applications for achieving short-term quantum advantage is \emph{Quantum Machine Learning (QML)}. This is mainly due to \emph{Variational Quantum Circuits (VQCs)} \cite{cerezo2021variational}.

VQCs are quantum circuits that depend on free parameters, which are iteratively updated by a classical optimizer to minimize an objective function estimated from measurements \cite{ostaszewski2021structure}. VQCs are especially suitable for NISQ devices, given their lower demands in the number of qubits and circuit depth, which mitigates the effect of noise \cite{cerezo2021variational}. Consequently, integrating VQCs into machine learning algorithms has become an intriguing field of study. VQCs used in such a context have been extensively researched in both supervised \cite{mitarai2018quantum, schuld2019quantum,schuld2020circuit,farhi2018classification} and unsupervised \cite{coyle2020born, zoufal2021variational} QML.

A RL agent learns from interaction with its surrounding environment in a feedback loop, relying solely on a reward signal to guide its actions. The concept of learning from interaction forms the bedrock of nearly all theories of learning and intelligence \cite{sutton2018reinforcement}, with some even conjecturing that rewards alone can pave the path to achieving general AI agents \cite{silver2021reward}. RL techniques, with the advent of deep learning \cite{lecun2015deep}, consecrated a major impact in tasks like Chess \cite{silver2017mastering}, Go \cite{silver2016mastering} and video-games \cite{mnih2015human}, matching or exceeding the performance levels of humans. It has also been applied to practical real-world problems, such as autonomous driving \cite{sallab2017deep}, robotics \cite{kober2013reinforcement} and finance \cite{hambly2023recent}.

Just like VQCs have found applications as function approximators in both supervised and unsupervised ML domains, they can be effectively utilized in the same capacity in RL scenarios, serving as alternatives to DNNs. These quantum variational approaches in the context of RL have started gaining traction only recently, as evidenced by several recent studies \cite{chen2020variational,lockwood2020reinforcement,skolik2022quantum,jerbi2021variational,sequeira2022variational, skolik2023equivariant}, which are further discussed on the next subsection.

\subsection{Related Work}

It was recently shown that the data encoding strategy used in the quantum circuit influences the expressive power of the quantum model \cite{schuld2021effect}. Specifically, the authors demonstrate that a VQC may be written as a Partial Fourier Series, where the accessible data frequencies are determined by the eigenvalues of the data-encoding Hamiltonians. Furthermore, quantum models can access increasingly rich frequency spectra by repeating simple data encoding gates multiple times either in series or in parallel, a technique named \emph{Data Re-uploading} \cite{perez2020data}.

VQCs have been used as function approximators in two different RL algorithms: Deep Q-learning \cite{chen2020variational,lockwood2020reinforcement,skolik2022quantum} and Reinforce with baseline \cite{jerbi2021variational,sequeira2022variational}. For the VQC-based Reinforce algorithms, \cite{jerbi2021variational} employed a VQC with data re-uploading and assessed the performance of models in benchmark environments. Conversely, \cite{sequeira2022variational} used a simpler, hardware-efficient ansatz without data re-uploading, comparing its performance and trainability against classical DNNs. Intriguingly, the impact of data re-uploading on model trainability remains unexplored. Given that data re-uploading increases the VQC's circuit depth and expressivity, it could have negative effects in the trainability, as suggested by \cite{mcclean2018barren,holmes2022connecting}.

The picture is even less clear when it comes to VQC-based Deep Q-Learning algorithms. Both \cite{chen2020variational} and \cite{lockwood2020reinforcement} used simple VQCs without data re-uploading, centering their studies on performance in benchmark environments and comparisons with DNNs. On the other hand, \cite{skolik2022quantum} integrated data re-uploading and analyzed its impact on performance. Yet, to the best of our knowledge, no research has been done on the trainability of VQC-based Deep Q-Learning models.

This paper aims to directly address these questions. Specifically, we seek to investigate the impact of data re-uploading on both the performance and trainability of VQC-based Deep Q-Learning models. Our approach involves empirically analyzing the average return of two different hardware-efficient VQCs in benchmark environments and the behavior of their gradient's norm and variance. Additionaly, we also verify how these metrics are affected by an increasing system size.

\subsection{Contributions}

The contributions of this paper are organized as follows:
\begin{itemize}
    \item Empirically verified the importance of data re-uploading and trainable output and input scaling in enhancing the performance of VQC-based Deep Q-learning models in OpenAI Gym's CartPole-v0 and Acrobot-v1.
    \item Achieved considerable performance in the Acrobot-v1 environment, which was previously untapped for VQC-based Deep Q-learning models.
    \item Empirically verified that the gradient's variance of the quantum models increases in the early stages of training in the aforementioned environments. Furthermore, we verified that the gradient's variance increases when data re-uploading is used.
    
    \item Empirically identified a correlation between the moving targets inherent to Deep Q-learning \cite{mnih2015human}, gradient magnitude and variance, and performance.
    \end{itemize}

\subsection{Outline}

The remainder of the manuscript is organized as follows. Section \ref{section:rl} introduces the mathematical framework behind most RL problems, the Q-Learning algorithm and the result of its combination with Deep Neural Networks. Section \ref{section:quantum} then introduces the main building blocks of Variational Quantum Circuits and how these can be used as function approximators in Deep Q-Learning. Section \ref{section:numerical} proposes the methodologies for analyzing performance and trainability and applies them to VQC-based Deep Q-Learning models in benchmark environments. Finally, \ref{section:conclusion} finishes the paper with some final remarks and suggestions for future work.

\section{Reinforcement Learning}\label{section:rl}
\subsection{Introduction}\label{subsection:introduction}

\begin{figure}[H]
    \centering
    \includegraphics[scale = 0.4]{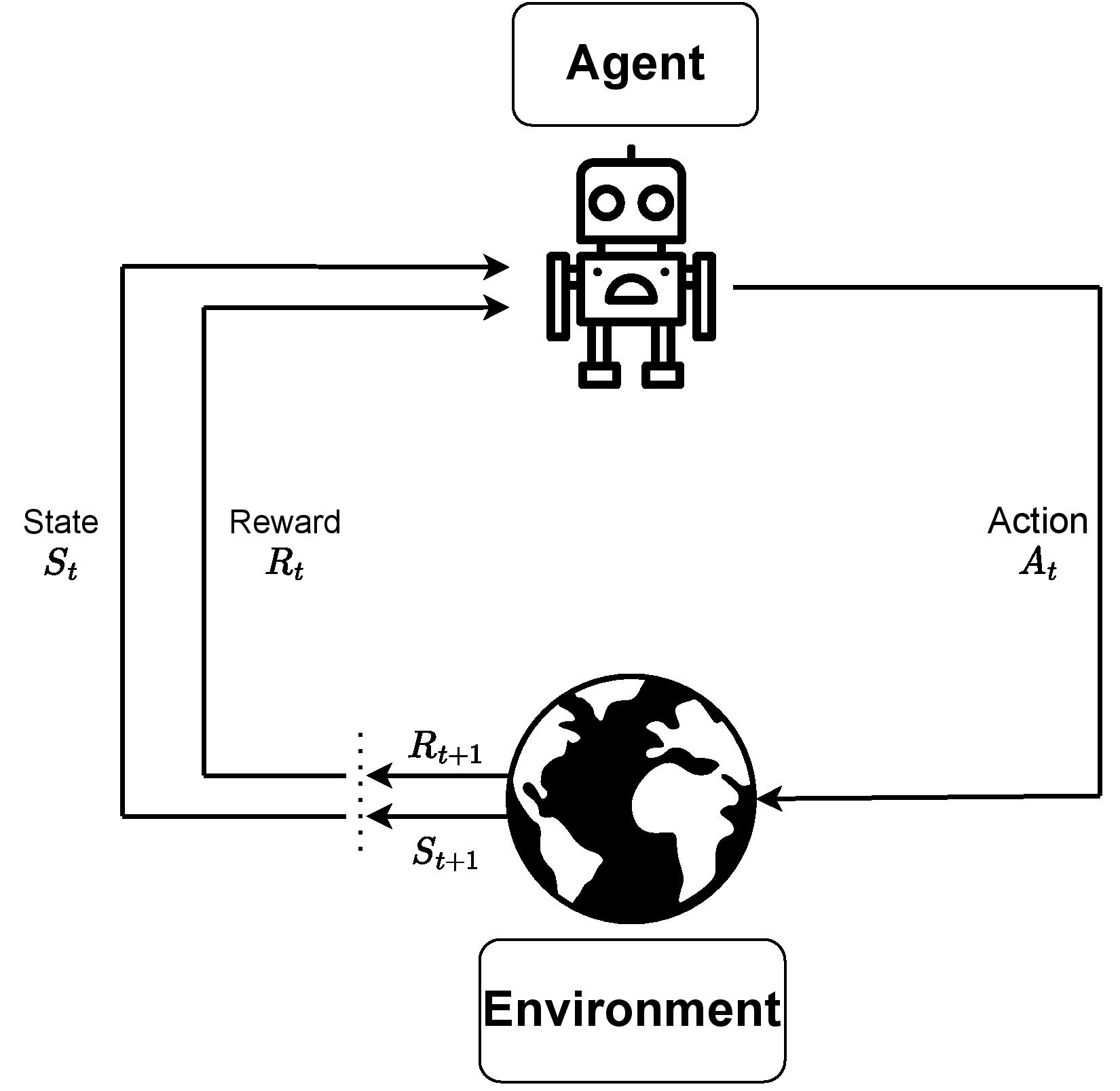}
    \caption{The Agent-Environment Interface: The agent interacts with the environment at time step $t$ by taking action $A_t$. The environment then changes to state $S_{t+1}$ and produces reward $R_{t+1}$, which are both passed back to the agent so that it can decide the next action. The dotted lines indicate that this process repeats itself. Inspired by \cite{sutton2018reinforcement}.}
    \label{fig:agent_environment_interface}
\end{figure}

Any RL problem has two primary constituents: the Agent and the Environment. The agent is the decision-maker and the representation of that which learns. The environment is everything that surrounds the agent and is affected by its decisions. The agent must learn which actions to take by interacting with the environment and receiving a reward that indicates the goodness of that particular action concerning the goal, see Figure \ref{fig:agent_environment_interface}.

The mathematical framework behind this interface and most RL problems is known as a \emph{Markov Decision Process (MDP)} \cite{sutton2018reinforcement}, which is expressed as a tuple $\langle \mathcal{S}, \mathcal{A}, \mathcal{P}, R\rangle$ \cite{sutton2018reinforcement}, encompassing a finite set of states $\mathcal{S}$, actions $\mathcal{A}$, a state transition probability matrix $\mathcal{P}$, and the reward function $R$.

A crucial metric in RL is the \emph{Return}, that is, the cumulative discounted reward starting from time-step $t$, such that:
\begin{equation}
    G_t = R_{t+1} + \gamma R_{t+2} + \gamma^2 R_{t+3}+... = \sum_{k=0}^{\infty}\gamma^{k}R_{t+k+1}
\end{equation}

where $\gamma \in [0,1]$ is the \emph{discount factor}, which determines how much long-term rewards are valued over short-term rewards. The behavior of the agent is given by the \emph{policy} $\pi(a|s)$, which is a probability distribution over actions given states. It determines the probability the agent will take each action $a$ from a particular state $s$. The objective of a RL agent becomes clear: to identify the policy that maximizes the expected return $\max_\pi\mathbb{E}_\pi[G_t]$. This policy is typically referred to as the \emph{optimal policy}. It is known that, for any MDP, there exists at least one deterministic optimal policy $\pi_*$ that is better than or equal to all other policies, $\pi_*\geq\pi,\forall\pi$ \cite{sutton2018reinforcement}. Consequently, the goal is for the policy of the agent to converge to this deterministic optimal policy as training goes on.

There are several methods for finding a (near-) optimal policy. We first distinguish between \emph{model-based} and \emph{model-free} methods \cite{sutton2018reinforcement,kaelbling1996reinforcement}. In model-based methods, the agent has access to (or learns an approximated version of) the model and can use it to plan. In model-free methods, the agent doesn't have access to a model. These methods are simpler and easier to implement, but also usually suffer from a higher sample complexity, since the agent has to deal with the extra burden of not knowing the dynamics of the environment.

Then, within the model-free methods, we need to distinguish between \emph{value-based} and \emph{policy-based} methods. The former use value-functions as guides in the search for optimal policies \cite{sutton2018reinforcement}, such as Q-learning \cite{watkins1992q}. Policy-based methods learn the policy directly. They typically work by parameterizing the policy $\pi_\theta(a|s)$, defining an objective function that serves as some performance measure $J(\pi_\theta)$ and optimizing the parameters $\theta$ by performing gradient ascent on an approximation of the gradient $\nabla J(\pi_\theta)$, which is obtained using Monte-Carlo samples. Consequently, they are commonly referred to as \emph{policy gradient} methods. The applicability of both value-based and policy-based methods vastly depends on the problem itself.

\subsection{Q-learning}\label{subsection:q_learning}

The \emph{Action-Value Function} $q_\pi(s,a)$ defines how good taking a certain action from a particular state is. It is defined as:
\begin{equation}
    q_\pi(s,a) = \mathbb{E}_\pi[G_t|S_t = s, A_t = a]
\end{equation}

Furthermore, in any MDP, there exists an optimal action-value function $q_*(s,a)$, that corresponds to the maximum action-value function over all policies
\begin{equation}
    q_*(s,a) = \max_\pi q_\pi(s,a)
\end{equation}
This function is achieved by all the optimal policies \cite{sutton2018reinforcement}. Actually, once the optimal action-value function is known, the optimal policy can be found by maximizing over it:
\begin{equation}\label{eq:maximizing}
    \pi_*(a|s) = \begin{cases}
        1 \;\;\text{if } a = \arg\max_{a\in\mathcal{A}}q_*(s,a)\\
        0 \;\;\text{otherwise}
    \end{cases}
\end{equation}

Q-Learning is a model-free value-based algorithm that approximates the optimal action-value function and then, according to Equation \ref{eq:maximizing}, a (near) optimal policy \cite{watkins1992q}.

Being model free, Q-Learning agents learn from experience - finite sequences of states, actions, and rewards, in the format $\{s,a,r,s^{\prime}\}$, which are called \emph{trajectories} \cite{sutton2018reinforcement}, or \emph{episodes} if the environment is episodic, which will be the case in this paper. To generate these episodes, some behavior policy has to be followed. Choosing this policy implies taking into consideration the \emph{exploration-exploitation tradeoff}. On the one hand, the agent has to explore the state space to find high-value states. On the other hand, the agent has to exploit the acquired knowledge, taking actions that it knows yield high rewards and lead to good states. As a result, typically, an \emph{$\epsilon$-greedy policy}, which chooses a random action with probability $\epsilon$ and the greedy action with probability $1-\epsilon$, is used \cite{sutton2018reinforcement}. It is usual to decay $\epsilon$ over time.

To approximate the optimal $Q$-values, Q-Learning uses a method known as \emph{Temporal-Difference Learning (TD-Learning)} - it learns from raw experience but updates estimates with other learning estimates without waiting for an outcome - \emph{bootstrapping}. More specifically, the action-value of a certain state and a particular action is updated as such:
\begin{equation}
    Q(S,A) \gets Q(S,A) + \alpha[R + \gamma \max_aQ(S^{\prime},a) - Q(S,A)]
\end{equation}

\subsection{Deep Q-Learning}\label{subsection:deep_qlearning}

The previous subsection introduces Q-Learning, which uses lookup tables of (at most) dimensions $\mathcal{S}\times\mathcal{A}$ to assign values to state-action pairs encountered by the agent - \emph{tabular regime}. Consequently, the algorithm is inadequate for handling large and complex real-world problems, since these tables become impossible to implement with current technology due to memory and complexity constraints.

There are several possible solutions for this problem, such as using state-abstraction \cite{andre2002state} and/or dimensionality reduction techniques \cite{sorzano2014survey}. This work focuses on the most popular solution - using \emph{Function Approximation}. The idea is to generalize from seen states to unseen states by approximating a target function. Consequently, a new function is defined:

\begin{equation}
    \hat{q}(s,a,\textbf{w}) \approx q_\pi(s,a)
\end{equation}

where $\textbf{w}$ is a vector of parameters that must be updated using TD-Learning or MC. It is important to refer that function approximation is only useful if $dim(\textbf{w})\ll |\mathcal{S}\times\mathcal{A}|$, otherwise it would be the same as the tabular regime.

There are many function approximators, each with pros and cons \cite{busoniu2017reinforcement}. Nonetheless, the most used function approximators are Neural Networks (NNs), particularly those with two or more hidden layers of neurons, known as Deep Neural Networks (DNNs), due to their capacity to approximate intricate and complex funtions \cite{lecun2015deep}. When DNNs are used as function approximators for the Q-function in a Q-Learning framework, the resultant algorithm is known as Deep Q-Learning or Deep Q-Networks (DQN) \cite{mnih2015human}. This algorithm introduces two novel improvements over previous approaches: \emph{experience replay} and \emph{target network}. The former consists of using a \emph{Replay Buffer}, where all experiences - interactions with the environment in the format $(s_t,a_t,r_{t+1},s_{t+1})$ - are stored. Then, batches of these experiences are randomly sampled to compute the parameter updates. Consequently, temporal correlations between the samples are broken, making the data appear more \emph{Independent and Identically Distributed (IID)}, which stabilizes training. The \emph{Target Network} consists of a second NN that is used to compute the TD-targets. The target network has "frozen" weights that are sporadically (every $C$ steps) updated with a copy of the parameters of the online network. Thus, the targets are more stationary, which also stabilizes training. However, there is a tradeoff here between speed of convergence and stability which must be accounted for when choosing $C$.

\begin{algorithm}
\caption{Deep Q-Learning \cite{mnih2015human}}
\label{DQN}
\begin{algorithmic}[1] 
\State Algorithm parameters: step size $\alpha\in[0,1]$, small $\epsilon\in [0,1]$
\State Initialize replay buffer $\mathcal{D}$ to capacity $N$
\State Initialize action-value function $Q$ with random weights $\theta$
\State Initialize target action-value function $\hat{Q}$ with weights $\theta^-=\theta$
\For{episode$=1,M$}
\State Initialize $s_1$
\For{$t=1,T$}
\State Choose $a_t$ from $s_t$ using $\epsilon$-greedy policy
\State Take action $a_t$, observe $r_{t+1}$, $s_{t+1}$
\State Store experience $(s_t,a_t,r_{t+1},s_{t+1})$ in replay memory $\mathcal{D}$
\State Sample random mini-batch of transitions $(s,a,r,s^{\prime})$ from $\mathcal{D}$
\State Compute the TD-targets using target network (with old, fixed parameters $\theta^-$)
\State Perform a gradient descent step on $\left[r + \gamma\max_{a^{\prime}}\hat{Q}(s^{\prime},a^{\prime};\theta_i^-) - Q(s,a;\theta_i)\right]^2$ w.r.t parameters $\theta$
\State Every $C$ steps update $\hat{Q} = Q$
\State Initialize $s_t$
\EndFor
\EndFor
\end{algorithmic}
\end{algorithm}

\section{Quantum Variational Deep Q-Learning}\label{section:quantum}

Quantum Variational Deep Q-learning, or VQC-based Deep Q-Learning, uses VQCs as function approximators in a Q-Learning framework, instead of the typically used DNNs. VQCs are quantum circuits that depend on free parameters, which are iteratively updated by a classical optimizer to minimize an objective function estimated from measurements \cite{ostaszewski2021structure}. In the following three subsections, we will go over the main details of the VQCs used throughout this work.

\subsection{Data Encoding}

Classical data has to be encoded into a quantum state so that it can be processed by a quantum computer. This paper uses \emph{Angle Encoding}. According to this technique, each component of the input data $x$ is encoded by a single qubit using arbitrary Pauli rotations ${R_X,R_Y,R_Z}$, where the angle is given by the component itself after some classical pre-processing, e.g. normalization. To allow for adaptive frequency matching between the function output by the VQC and the target function, each feature will be multiplied by a classical trainable weight, a technique known as trainable input scaling \cite{perez2020data}. Thus, given the input data $x = \{x_0,x_1,...,x_{n-1}\}$, the resulting quantum state is given by:
\begin{equation}
    \ket{x} = \bigotimes_{i=0}^{n-1}R_\alpha(\phi(x_i \times \lambda_i))\ket{0_i}
\end{equation}

where $R_\alpha\in\{R_X,R_Y,R_Z\}$, $\phi$ is some classical pre-processing function and $\ket{0_i}$ is the $i^{\text{th}}$ qubit initialized in state $\ket{0}$.

This technique has pros and cons. On the one hand, it allows the encoding of a given input vector using circuits of depth $1$ \cite{sequeira2022variational}. On the other hand, the number of qubits grows linearly with the number of features of the input vector, which limits the dimensionality of the inputs one can encode due to the reduced number of qubits in NISQ devices.

\subsection{Ansatz}

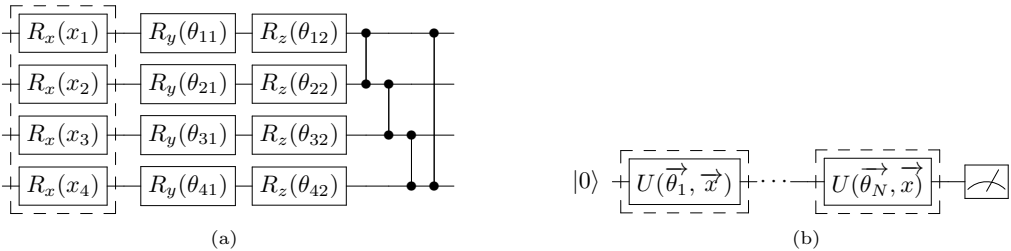
\begin{figure}[H]
     \centering
     \begin{subfigure}[b]{0.5\textwidth}
         \centering
         \scalebox{0.9}{
                \Qcircuit @C=0.7em @R=.5em {
                    & \gate{R_x(x_1)} & \qw & \gate{R_y(\theta_{11})} & \gate{R_z(\theta_{12})} & \ctrl{1} & \qw & \qw & \ctrl{3} & \qw\\
                    & \gate{R_x(x_2)} & \qw & \gate{R_y(\theta_{21})} & \gate{R_z(\theta_{22})} & \ctrl{-1} & \ctrl{1} & \qw & \qw & \qw\\
                    & \gate{R_x(x_3)} & \qw & \gate{R_y(\theta_{31})} & \gate{R_z(\theta_{32})} & \qw & \ctrl{-1} & \ctrl{1} & \qw & \qw\\
                    & \gate{R_x(x_4)} & \qw & \gate{R_y(\theta_{41})} & \gate{R_z(\theta_{42})} & \qw & \qw & \ctrl{-1} & \ctrl{-3} & \qw \gategroup{1}{2}{4}{2}{.7em}{--}\\
                }
                }
         \caption{}
         \label{fig:skolik}
     \end{subfigure}%
     \begin{subfigure}[b]{0.5\textwidth}
         \centering
         \scalebox{0.9}{
    \Qcircuit @C=0.7em @R=.5em {
        \lstick{\ket{0}} & \gate{U(\overrightarrow{\theta_1}, \overrightarrow{x})} & \qw & \cdots & & \qw & \gate{U(\overrightarrow{\theta_N}, \overrightarrow{x)}} & \qw & \meter \gategroup{1}{2}{1}{2}{.7em}{--} \gategroup{1}{7}{1}{7}{.7em}{--}
            }
        }
         \caption{}
         \label{fig:uqc}
     \end{subfigure}
     \caption{Subfigure \ref{fig:skolik}: Skolik's Architecture. When Data Re-Uploading is used, the whole circuit is repeated in each layer. Otherwise, just the part that is not surrounded by dashed lines. Subfigure \ref{fig:uqc}: UQC Architecture. Each processing layer $U$ is given by $U^{UAT}(\overrightarrow{x};\overrightarrow{\omega},\alpha,\varphi) = R_y(2\varphi)R_z(2\overrightarrow{\omega}\cdot\overrightarrow{x}+2\alpha)$ and $\overrightarrow{\theta}_i = (\overrightarrow{\omega} , \alpha, \varphi)$. Although a single-qubit ansatz was shown for simplicity, this ansatz can be generalized to allow multiple qubits.}
     \label{fig:ansatzes}
\end{figure}

A set of problem-agnostic ansatzes called \emph{Hardware-Efficient Ansatzes} were considered in this work. These allow for implementations with reduced circuit depth by bringing correlated qubits together for depth-reduction \cite{cerezo2021variational}. For this work, the ansatzes selected are those from Skolik’s \cite{skolik2022quantum} and the Universal Quantum Classifier (UQC) \cite{perez2020data}, depicted in Figure \ref{fig:ansatzes}. For simplicity, the Skolik's architecture will be referred to as \emph{Skolik Data Re-Uploading} if data re-uploading is used and \emph{Skolik Baseline} otherwise.

However, there is a tradeoff one needs to take into account when using hardware-efficient VQCs. Since these ansatzes take no inspiration from the structure of the problem itself, they need to be highly expressive so they can be applied to any generic task \cite{bilkis2021semi}. However, this expressiveness comes at the cost of reduced trainability \cite{holmes2022connecting}. This problem, known as the \emph{Barren-Plateau Phenomenon}, is explained in Subsection \ref{subsection:BP}.

\subsection{Cost Function and Gradient-Based Learning}

To use a VQC to solve a problem, one has to first encode the problem into a cost function, such that solving the problem corresponds to finding the global optima of said function. Let $f_{\boldsymbol{\theta}}(x)$ be the expectation value of some observable $\hat{O}$ as follows:
\begin{equation}\label{eq:expectation}
    f_{\boldsymbol{\theta}}(x) = \bra{0}U^\dagger(x,\boldsymbol{\theta})\hat{O}U(x,\boldsymbol{\theta})\ket{0}
\end{equation}
where $U(x,\boldsymbol{\theta})$ is a VQC that depends on the input $x$ and a set of free parameters $\boldsymbol{\theta}$.

In the context of VQC-based Q-learning, considering an environment with $|A|$ possible actions, we need to measure the expectation values of $|A|$ observables, such that the Q-value of a particular action $a$ is given by:

\begin{equation}
    Q(s,a) = \bra{0^{\otimes n}}U(s,{\boldsymbol{\theta}})^\dagger O_aU(s,{\boldsymbol{\theta}})\ket{0^{\otimes n}}
\end{equation}

where $s$ is the state, $n$ the number of qubits and $O_a$ the observable corresponding to action $a$.

The observables used throughout this work depend on the ansatz and number of qubits, but always consist of single-qubit $\sigma_z$'s or tensor products of $\sigma_z$'s applied to different qubits, similar to what has been previously used in the literature \cite{skolik2022quantum}. 

Returning to the general scenario of a function $f_{\boldsymbol{\theta}}(x)$, the cost function $L(\boldsymbol{\theta})$ is a function of the expectation value itself. In many scenarios, $L(\boldsymbol{\theta})$ reduces to $f_{\boldsymbol{\theta}}(x)$. However, in the context of machine learning, classical post-processing is usually added to the expectation value. For instance, consider the typical supervised learning scenario. Given a labeled dataset $D = \{(x_i,y_i)\}^{M}$, the typical cost function is the Mean Squared Error (MSE), which can be expressed as follows:
\begin{equation}
    L(\boldsymbol{\theta}) = \frac{1}{M}\sum_{i=0}^{M-1}\left(f_{\boldsymbol{\theta}}(x_i) - y_i\right)^{2}
\end{equation}

In practice, the value of $f_{\boldsymbol{\theta}}$ for a specific input $x$ is given by running the VQC multiple times and averaging over the results. Then, typically, gradient-based methods are used to update the parameters of the VQC. However, gradient-free methods also exist and have their advantages and disadvantages, see \cite{cerezo2021variational}.

When using gradient-based methods, as will be the case of this work, the gradient of the cost function with respect to the parameters is calculated using the Parameter-Shift Rules \cite{schuld2019evaluating}. These rules state that the partial derivative of $f_{\boldsymbol{\theta}}(x)$ w.r.t a single variational parameter (assuming this parameter is the angle of a Pauli-rotation) is given by:
\begin{equation}
\frac{\partial f_{\boldsymbol{\theta}}(x)}{\partial \theta_i} = \frac{1}{2} \left[ f_{\boldsymbol{\theta}}(x; \theta_i + \pi/2) - f_{\boldsymbol{\theta}}(x; \theta_i - \pi/2) \right]
\end{equation}
To compute the partial derivative of the cost function w.r.t a single parameter, two circuit executions are needed. Thus, to compute the gradient of the cost function, which has $p$ parameters, $2\times p$ executions are necessary. However, it is important to note that, in the numerical experiments of this paper, quantum simulators and the adjoint differentiation techniques will be used \cite{jones2020efficient}.

\subsection{VQC's Trainability}\label{subsection:BP}

Let $L(\boldsymbol{\theta})$ be a hardware-efficient VQC cost-function. The gradient of this function is given by:
\begin{equation}\label{eq:gradient}
    \nabla_{\boldsymbol{\theta}} L(\boldsymbol{\theta}) = \left\{\frac{\partial L}{\partial \theta_1},\frac{\partial L}{\partial \theta_2},...,\frac{\partial L}{\partial \theta_k}\right\},\;\boldsymbol{\theta}\in \mathbb{R}^k
\end{equation}
where $\boldsymbol{\theta}$ are the cost function's parameters and $k$ the number of parameters.

The Barren Plateau Phenomenon asserts that
\begin{equation}\label{eq:barren_plateau}
Var\left( \frac{\partial L(\boldsymbol{\theta})}{\partial \theta_k}\right)\in\mathcal{O}\left(\frac{1}{\alpha^{n}}\right),\;\alpha>1
\end{equation}
where $n$ is the number of qubits.

Put in words, the variance of the gradient of hardware-efficient VQC-based cost functions decays exponentially with the number of qubits \cite{mcclean2018barren}. Furthermore, \cite{holmes2022connecting} demonstrated that the more expressive the ansatz, the lower the variance in the cost gradient. Consequently, the cost landscape is flatter, making these circuits extremely hard to train. Nonetheless, it is important to emphasize that \cite{cerezo2021cost} observed that the Barren Plateau Phenomenon highly depends on the locality of the observable employed. To mitigate this issue, several methods have been proposed, such as Barren-Plateau Free ansatzes \cite{park2024hamiltonian,schatzki2024theoretical}. However, Cerezo et al. \cite{cerezo2023does} claim that circuits that can be proved to be free from Barren Plateaus actually end up encoding the problem in a polynomial sub-space of the Hilbert Space, one that is classically simulatable. 

It is important to note that the BP Phenomenon is a statement over the landscape on average \cite{drudis2024variational}, thus it does not exclude the possibility of there being fertile valleys - regions with high gradients near good local minima. Reasearch on warm starts aims to initialize the VQCs in such regions of the landscape \cite{mele2022avoiding,zhang2022escaping, grant2019initialization}. Care must be taken, however, since regions with large gradients are not necessarily near a good local minima. In fact, it has been shown that hardware efficient VQCs’ landscapes are swamped with bad local minima [46].

\subsection{Data Re-Uploading}

The authors of \cite{schuld2021effect} determined that the data encoding strategy used in the quantum circuit influences the expressive power of the quantum model. Specifically, they show that a VQC may be written as a Partial Fourier Series, where the accessible data frequencies are determined by the eigenvalues of the data-encoding Hamiltonians. Furthermore, quantum models can access increasingly rich frequency spectra by repeating simple data encoding gates multiple times either in series or in parallel, a technique called \emph{Data Re-uploading} \cite{perez2020data}.

To exemplify the power of data re-uploading, the authors showed that a quantum model employing a single Pauli-rotation encoding can only learn a sine function of the input. In other words, if the encoding is done just once, the VQC can only approximate functions with a single non-zero frequency in the frequency spectrum. However, repeating Pauli encoding linearly extends the frequency spectrum, such that, if the encoding is repeated $n$ times, then the quantum model can approximate functions with up to $n$ different non-zero frequencies.

\subsection{VQC-based Deep Q-Learning}

Just like NNs can be used to approximate either a value-function or a parameterized policy, so can VQCs. The result is a hybrid classical-quantum algorithm that generally works as follows. The agent observes some state $s_t$ and applies some classical pre-processing $\phi$. Then, the result $\phi(s)$ is encoded into a VQC $U_{\boldsymbol{\theta}}(\phi(s))$ using some data encoding technique. The VQC, with the current parameters $\boldsymbol{\theta}_t$, prepares a quantum state. An observable $O_a$ is measured for each possible action. The expectation values of these observables $\expval{O_a}_{s,\boldsymbol{\theta}}$ are post-processed and the result represents either the Q-values $Q_{\boldsymbol{\theta}}(s,a)$ in value-based methods or the policy $\pi_{\boldsymbol{\theta}}(a|s)$ in policy-based methods. Then, the agent chooses an action $a_t$ using these predictions and executes it in the environment. The reward $r_t$ and the consecutive state $s_{t+1}$ are observed by the classical optimizer. Using the parameter-shift rule, the gradients of the VQC w.r.t the parameters $\boldsymbol{\theta}_t$ are calculated. Finally, the classical optimizer determines the new parameters $\boldsymbol{\theta}_{t+1}$ \cite{meyer2022survey}.

In VQC-based Deep Q-Learning, in order for the Q-values estimated by the quantum circuit to match the optimal Q-values in the environment, the expectation values of the observables will be multiplied by a classical trainable weight, a technique known as \emph{Trainable Output Scaling} \cite{skolik2022quantum}.

To the best of our knowledge, \cite{chen2020variational} was the first paper to use VQCs as function approximators in RL algorithms. Specifically, the authors used VQCs to approximate the Q-function in a Deep Q-Learning algorithm. The pseudo-code for the algorithm can be seen in Algorithm \ref{Variational Quantum Deep Q-Learning}.

\begin{algorithm}[h]
\caption{Variational Quantum Deep Q-Learning}
\label{Variational Quantum Deep Q-Learning}
\begin{algorithmic}[1] 
\State Algorithm parameters: step size $\alpha\in[0,1]$, small $\epsilon\in [0,1]$
\State Initialize replay buffer $\mathcal{D}$ to capacity $N$
\State Initialize action-value function $Q$ (quantum circuit) with random parameters $\theta$
\State Initialize target action-value function $\hat{Q}$ (target quantum circuit) with parameters $\theta^-=\theta$
\For{episode$=1,M$}
\State Initialize $s_1$ and encode into the quantum state
\For{$t=1,T$}
\State With probability $\epsilon$ select a random action $a_t$, otherwise select $a_t = \max_a Q^*(s_t,a;\theta)$ from the output of the quantum circuit
\State Execute action $a_t$ in emulator and observe $r_{t+1}$, $s_{t+1}$
\State Store experience $(s_t,a_t,r_{t+1},s_{t+1})$ in replay memory $\mathcal{D}$
\State Sample random mini-batch of transitions $(s,a,r,s^{\prime})$ from $\mathcal{D}$
\State Compute the TD-targets using target quantum circuit (with old, fixed parameters $\theta^-$)
\State Perform a gradient descent step on $\left[r + \gamma\max_{a^{\prime}}\hat{Q}(s^{\prime},a^{\prime};\theta_i^-) - Q(s,a;\theta_i)\right]^2$ w.r.t parameters $\theta$
\State Every $C$ steps update $\hat{Q} = Q$
\State Initialize $s_t$ and encode into the quantum state
\EndFor
\EndFor
\end{algorithmic}
\end{algorithm}

\section{Numerical Results}\label{section:numerical}

In this section, we present and analyse experimental numerical results concerning both the performance and trainability of VQC-based DQN agents. The results are summarized in table \ref{table:numerical_results}.

\begin{table}[h]
\caption{}
\label{table:numerical_results}
\begin{tabular}{@{}p{0.2\linewidth} p{0.4\linewidth} p{0.4\linewidth}@{}}
\toprule
\textbf{Figure} & \textbf{Explanation} & \textbf{Conclusions} \\
\midrule
Fig. \ref{fig:skolik_performance} & Analysis of the effect data re-uploading and input/output scaling have on the performance of QRL agents in CartPole-v0 and Acrobot-v1 environments. & Data re-uploading, input scaling, and output scaling all increase the agents' performance. \\
Fig. \ref{fig:skolik_trainability} & Analysis of the trainability of the agents from Fig \ref{fig:skolik_performance}. & Gradients tend to increase throughout training and Data Re-uploading models exhibit higher gradients than baseline models.\\
Figs. \ref{fig:losses_target_network} and \ref{fig:gradients_target_network} & Analysis of the loss and its gradients throughout training for different values of the target network update frequency. & It's the instability that arises from low target network update frequency values that leads to the gradients' behavior observed in Fig. \ref{fig:skolik_trainability}.\\
Fig. \ref{fig:uqc_entanglement} & Performance of the Single and Multi-Qubit UQC with and without entanglement on the Cartpole-v0 and Acrobot-v1 environments. & When no entanglement is used, all the features of the state must be encoded into all qubits, otherwise the agent's lack the necessary information to achieve (near) optimal policies.\\
Fig. \ref{fig:variance_first_step} & Variance of the gradient for $1000$ uniformly initialized UQC agents with local and global cost functions as the number of qubits increases. & The variance of the gradient decays exponentially when a global cost function is used and in the border of a polynomial to exponential regime when a local cost function is used.\\
Fig. \ref{fig:number_of_qubits} & Performance and norm/variance of the gradients throughout training of the UQC models for an increasing number of qubits. & The gradients exhibit a similar behavior to what was previously seen. Moreover, as the number of qubits increases, these gradients remain high.\\
Figs. \ref{fig:supervised_performance} and \ref{fig:supervised_gradients} & Performance and Gradients of the UQC and Skolik models trained on a binary classification dataset for an increasing number of qubits. & The UQC model consistently achieves higher accuracies than the Skolik models. Moreover, these accuracies do not decrease as the number of qubits increases. Finally, the gradients do not exhibit the same behavior as seen in Fig \ref{fig:number_of_qubits}. Thus, the moving targets of Deep Q-Learning contribute to the large gradients found throughout training.\\

\bottomrule
\end{tabular}
\end{table}

For the Skolik models, the input and output scaling weights were initialized as $1$s while the rotational parameters $\theta \sim U[0,\pi]$. For the UQC models, the weights $\Vec{w}\sim \mathcal{N}(0,0.01)$ , the bias $\Vec{b}$ were initialized as $0$s, the rotational parameters $\varphi\sim U[0,\pi]$ and the output scaling weights as $1$s. For the full hyperparameters for each experiment, see \ref{secA1}.

\subsection{Methodology for Analysing Performance and Trainability}

In any RL problem, the goal is for the agent to find the optimal or a near-optimal policy, which translates to the behaviour that accumulates the most rewards. Consequently, performance is typically measured by the return models achieve in benchmark environments.

To analyze performance, we use the following methodology: $N$ agents are randomly initialized according to the model. These agents are then trained in the environment over $M$ episodes. Should an agent solve the environment before completing the $M$ episodes, its training ceases (i.e., no further updates to the model's parameters), but the agent continues interacting with the environment using its learned knowledge until all episodes conclude. The returns from each episode are collected for all agents. Ultimately, we average the returns across all $N$ agents for each episode, compute the standard deviation, and plot the results. The best-performing model will be the one that achieves the highest average return in less episodes.

To analyse trainability, the following methodology is used: when analysing a model's performance, we also store the gradient of the cost function with respect to the parameters at each training step, which is every step where the parameters are updated. Subsequently, we compute the norm of these gradients for every training step. To mitigate the effects of stochasticity, we calculate the average norm of the gradients across the $N$ agents and the variance of these norms at each training step. However, since agents cease training once the environment is solved, different agents have a different number of training steps. As a result, we limited our procedure to only encompass the steps up until the first agent successfully solved the environment. To facilitate the visualization of the results, a rolling average of the last $100$ training steps is used for both the norm of the gradients and the variance of the norms.

\subsection{Environments}

OpenAI Gym, an open-source Python library, offers a standardized API that bridges learning algorithms with a standard set of environments, providing a platform to analyse and compare RL algorithms. Two specific environments were chosen: CartPole-v0 and Acrobot-V1. The preference for these is twofold: firstly, they are established benchmark environments that have been frequently adopted in academic research for algorithm testing and comparison. Secondly, these environments strike a balance in complexity; they present sufficient challenges to test the robustness of VQC-based algorithms, yet remain tractable for experimental purposes. While CartPole has already been the subject of investigation in several studies, Acrobot presents a heightened degree of complexity and, intriguingly, hasn't yet been solved using VQC-based Q-Learning.

CartPole consists of a cart that moves on a frictionless track with a pendulum placed upright. The goal is to balance the pole by applying forces in the left and right direction on the cart. The state-space is composed of $4$ features that include the cart's position and velocity and the pole's angle and angular velocity. The reward function is very simple: the agent receives a $+1$ reward per time-step and there are a maximum of $200$ time-steps.

Acrobot consists of two links that are linearly connected to form a chain, with one end of the chain fixed. The joint between the two links is the one we can act upon and the goal is to apply torques ($-1$,$0$ or $1$) to this joint to swing the free end of the chain above a given height while starting from the initial position of hanging downwards. The state-space is more complex than Cartpole's and consists of $6$ features. Nonetheless, we employed a trivial technique to reduce it to $4$. Every step that does not reach the goal receives a reward of $-1$ and achieving it results in termination with a reward of $0$. Moreover, the maximum number of steps in an episode is $500$, after which the episode also terminates.

\subsection{Data Re-Uploading's Effect on Performance}\label{subsection:skolik_performance}

\begin{figure}[h]
    \centering
    \begin{subfigure}[t]{0.9\textwidth}
        \includegraphics[width=\textwidth]{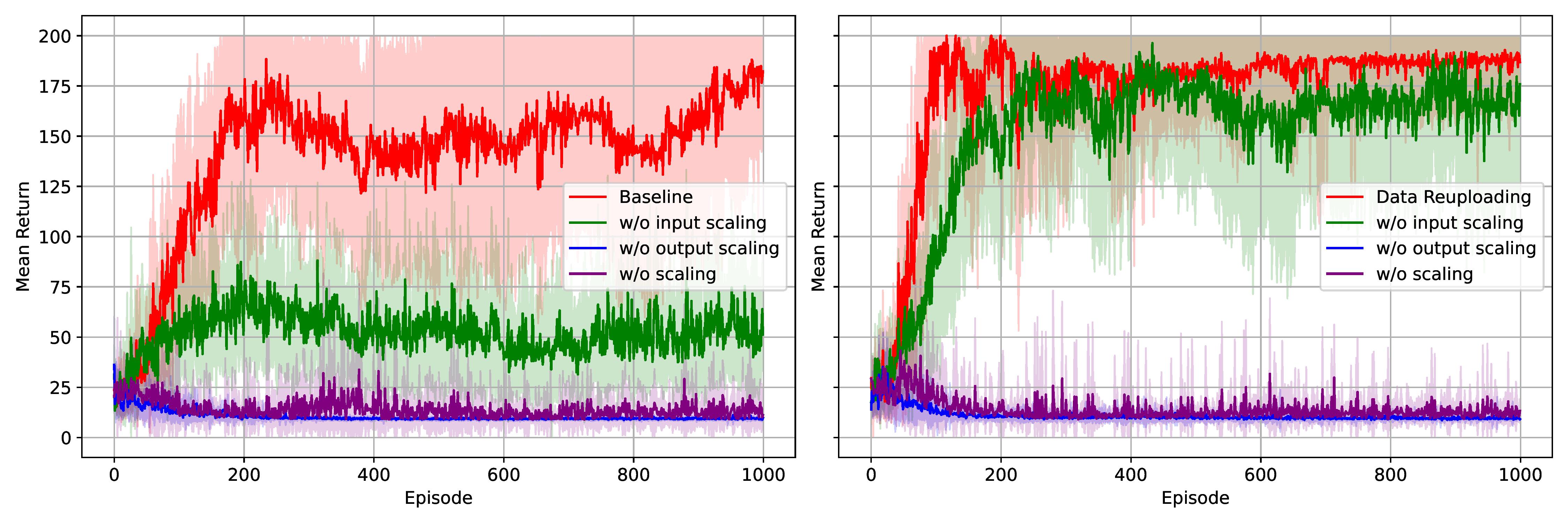}
        \caption{}
        \label{fig:skolik_cartpole}
    \end{subfigure}
    \begin{subfigure}[t]{0.9\textwidth}
        \includegraphics[width=\textwidth]{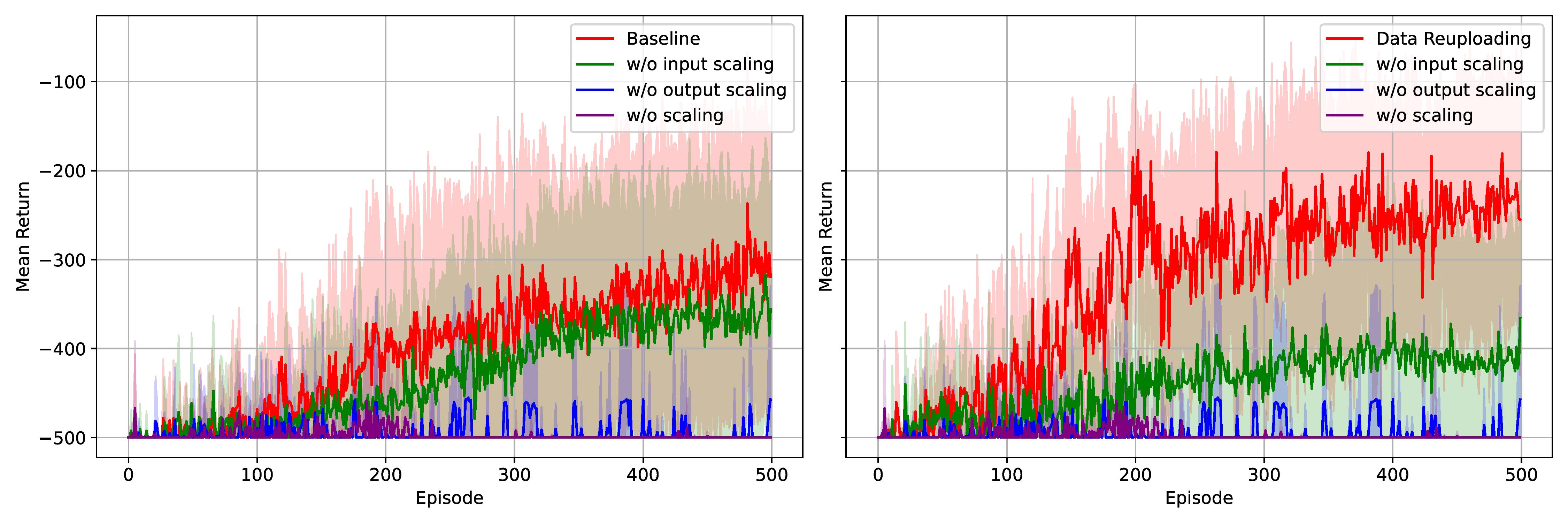}
        \caption{}
        \label{fig:skolik_acrobot}
    \end{subfigure}
    \caption{Performance of Baseline Models (on the left) and Data Re-Uploading models (on the right) in the CartPole-v0 environment (see Subfigure \ref{fig:skolik_cartpole}) and the Acrobot-v1 environment (see Subfigure \ref{fig:skolik_acrobot}) with and without trainable input and/or output scaling. The returns are averaged over $10$ agents. The full set of hyperparameters can be seen in Table \ref{table: skolik_hyper}.}
    \label{fig:skolik_performance}
\end{figure}

To illustrate the impact of data re-uploading and trainable input and output scaling, we analyzed the performance of models composed of all the possible combinations of these characteristics, see Figure \ref{fig:skolik_performance}.

Figures \ref{fig:skolik_cartpole} and \ref{fig:skolik_acrobot} show the considerable effects of data re-uploading and trainable input and output scaling on the performance of VQC-based Deep Q-Learning agents in the CartPole-v0 and Acrobot-v1 environments. Agents that do not use trainable output scaling perform very poorly, no better than random guessing, demonstrating the importance of matching the range of the Q-values estimated by the quantum circuit to the range of the optimal Q-values of the specific environment. While it is possible to achieve a considerable level of performance by multiplying the expectation values by a fixed value instead of using a classical trainable weight, \cite{skolik2022quantum} show that such a technique depends on the speed of convergence of the models and does not achieve the same level of performance as models that use trainable output scaling.

Moreover, the impact of trainable input scaling is noticeable, considering that models that use it consistently outperform models that do not. This result emphasizes the importance of allowing for an adaptive frequency matching between the function the VQC outputs and the one it attempts to approximate \cite{perez2020data}. 

Finally, most models that use data re-uploading outperform those that do not in both environments, which reaffirms the results obtained by \cite{skolik2022quantum}, demonstrating the importance of such a technique in increasing the expressivity of the VQC and, consequently, allowing for a better approximation of the optimal Q-function. However, this is not the case for the models without trainable input scaling in the Acrobot-v1 environment, which is surprising. This result may possibly be explained by the statistical variance inherent to our methodology for analysing performance. Nonetheless, if the trainability of data re-uploading models is considerably reduced, then the applicability of data re-uploading might be severely limited. Hence, the next section analyses the trainability of these models.

\subsection{Data Re-Uploading's Effect on Trainability}

\begin{figure}[h]
    \centering
    \begin{subfigure}[t]{0.9\textwidth}
        \includegraphics[width=\textwidth]{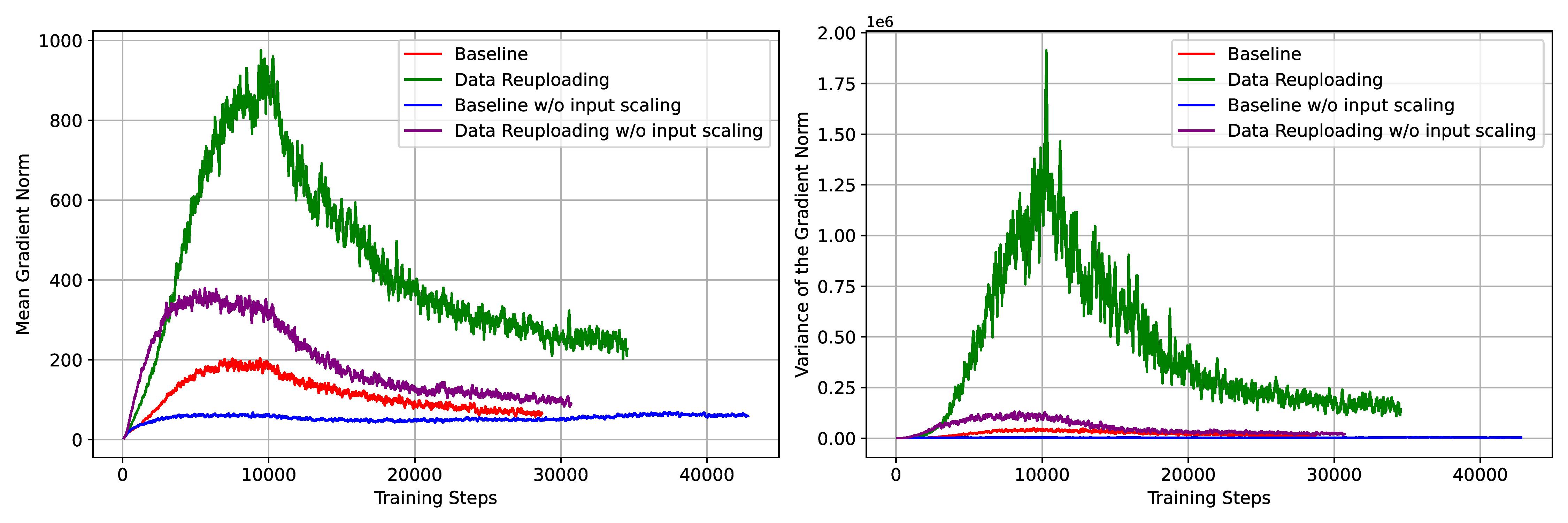}
        \caption{}
        \label{fig:skolik_gradients_cartpole}
    \end{subfigure}
    \begin{subfigure}[t]{0.9\textwidth}
        \includegraphics[width=\textwidth]{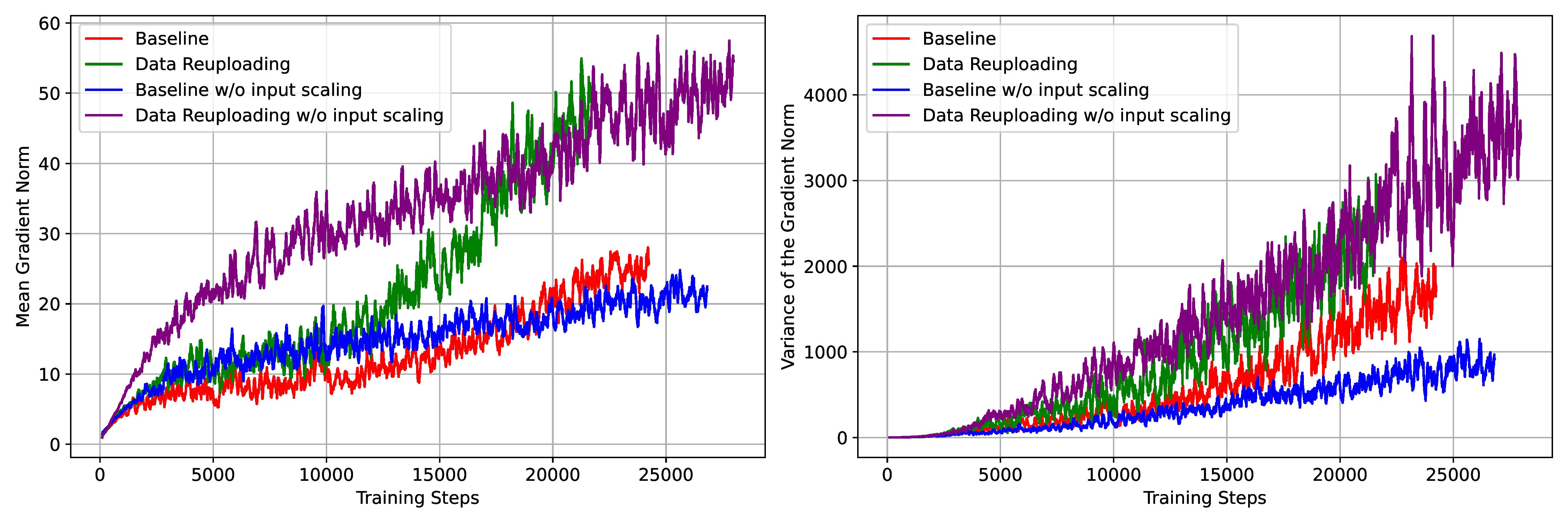}
        \caption{}
        \label{fig:skolik_gradients_acrobot}
    \end{subfigure}
    \caption{Trainability of the models from Figures \ref{fig:skolik_cartpole} and \ref{fig:skolik_acrobot} with trainable output scaling in the CartPole-v0 (see Subfigure \ref{fig:skolik_gradients_cartpole}) and Acrobot-v1 (see Subfigure \ref{fig:skolik_gradients_acrobot}) environments. In both Subfigures, the left graph represents the gradient's norm throughout training and the right graph the variance of the norm.}
    \label{fig:skolik_trainability}
\end{figure}

In this section, we turn our attentions to the trainability of the models from Figures \ref{fig:skolik_cartpole} and \ref{fig:skolik_acrobot}. Since the models without trainable output scaling achieved a poor level of performance, they were excluded from this analysis. The results can be seen in Figures \ref{fig:skolik_gradients_cartpole} and \ref{fig:skolik_gradients_acrobot}.

These results are noteworthy in several ways. First, there is a pattern in the gradient's norm and variance. From Subfigure \ref{fig:skolik_gradients_cartpole}, one can see that they increase in the training's early stages, peak and then decrease on Cartpole-v0. Interestingly, they increase when the agent needs to learn the most (when it is initialized and the environment is new) and start to decrease around the $10000$ training-step mark when the agent should have done most of the learning already and started converging to a policy. However, the gradients of the models trained on Acrobot-v1 consistently increase throughout training, see Subfigure \ref{fig:skolik_gradients_acrobot}. This could possibly be explained by the fact that the agents never reached a consistently high return in this environment.

Moreover, despite their greater circuit depth and enhanced expressivity, the data re-uploading models exhibit the highest gradient's variance among all models. This findind is in stark contrast to our initial expectations derived from \cite{holmes2022connecting}, which suggests a trade-off between expressivity and trainability. We should nonetheless note that \cite{holmes2022connecting} claims that more expressive models are harder to train in the context of Barren Plateaus, that is, the gradient's variance of such models decreases exponentially with system size upon initialization (when parameters are initialized uniformly spanning the whole Hilbert Space). Thus, our results and theirs are not mutually exclusive in the sense that our models may still suffer from the same problem upon uniform initialization covering the whole Hilbert Space, but our initialization in the context of Deep Q-learning leads to large gradients throughout training, a similar concept to warm-starts.

\subsection{Trade-off Between Moving Targets and Gradient Magnitude}\label{subsection:tradeoff}

\begin{figure}[h]
    \centering
    \begin{subfigure}[t]{0.4\textwidth}
        \includegraphics[width=\textwidth]{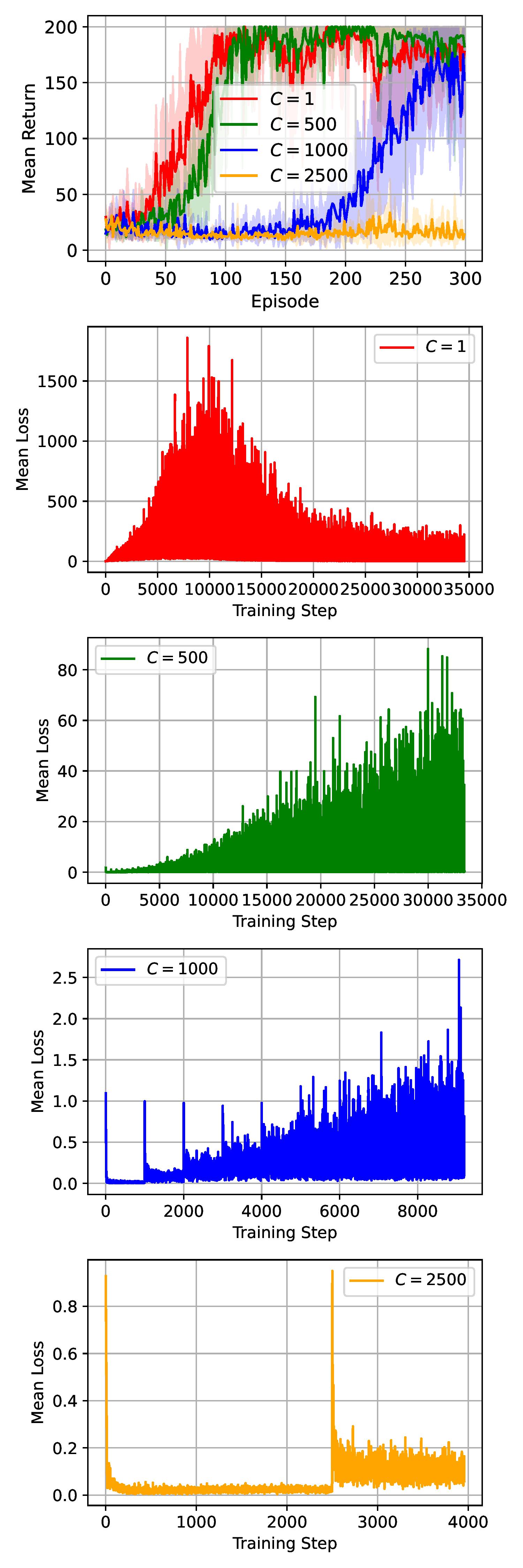}
        \caption{}
        \label{fig:target_loss_cartpole}
    \end{subfigure}
    \begin{subfigure}[t]{0.4\textwidth}
        \includegraphics[width=\textwidth]{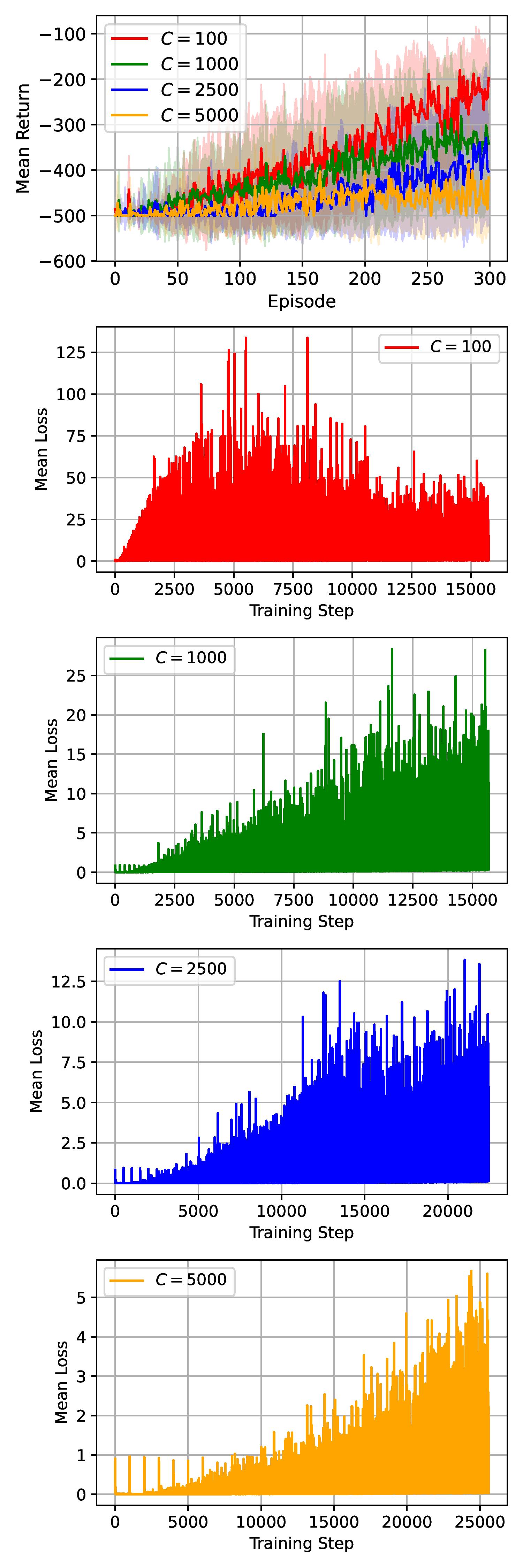}
        \caption{}
        \label{fig:target_loss_acrobot}
    \end{subfigure}
    \caption{Performance (first graph) and the respective loss functions for increasing values of $C$ of data re-uploading models in the Cartpole-v0 (see Subfigure \ref{fig:target_loss_cartpole}) and Acrobot-v1 (see Subfigure \ref{fig:target_loss_acrobot}) environments. The full set of hyperparameters can be seen in Table \ref{table:target_loss}.}
    \label{fig:losses_target_network}
\end{figure}

In this section, we attempt to comprehend why the norm and the variance of the gradients exhibit such a surprising behavior, both in the sense that they reach substantial maximum values and in the sense that more expressive models achieve higher maximum values. For a start, it is crucial to understand that Deep Q-Learning is fundamentally different from supervised learning, with the primary reason being that Deep Q-Learning targets are \emph{non-stationary}. Since the targets keep changing during training due to the agent's evolving knowledge, predicting Q-values becomes increasingly challenging. This is specially pronounced in the beginning of training, when the agent is focused on exploring the state-space. The more states explored, the higher the variance in the return and, consequently, the higher the loss.

Due to the instability that arose from the moving targets, the original paper \cite{mnih2015human} introduced the concept of a target network. The targets are calculated using this network that has frozen weights, which are updated every $C$ steps to match the weights of the online network. Hence, for $C$ steps at a time, the targets appear stationary. If $C$ is set too high, the targets move slowly and the model takes longer to train. If $C$ is too low, then the targets change frequently and the algorithm becomes unstable.

Figures \ref{fig:target_loss_cartpole} and \ref{fig:target_loss_acrobot} show how data re-uploading models perform for different values of $C$ and the behavior of their loss functions in the CartPole-v0 and Acrobot-v1 environments. 

\begin{figure}[h]
    \centering
    \begin{subfigure}[t]{\textwidth}
        \includegraphics[width=\textwidth]{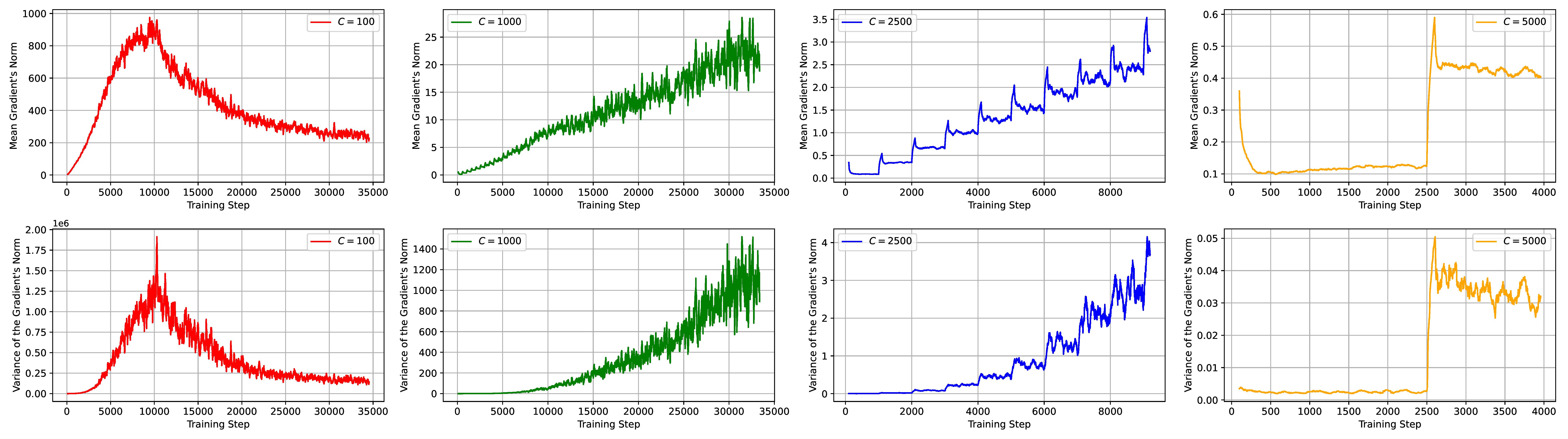}
        \caption{}
        \label{fig:gradients_behavior_cartpole}
    \end{subfigure}
    \begin{subfigure}[t]{\textwidth}
        \includegraphics[width=\textwidth]{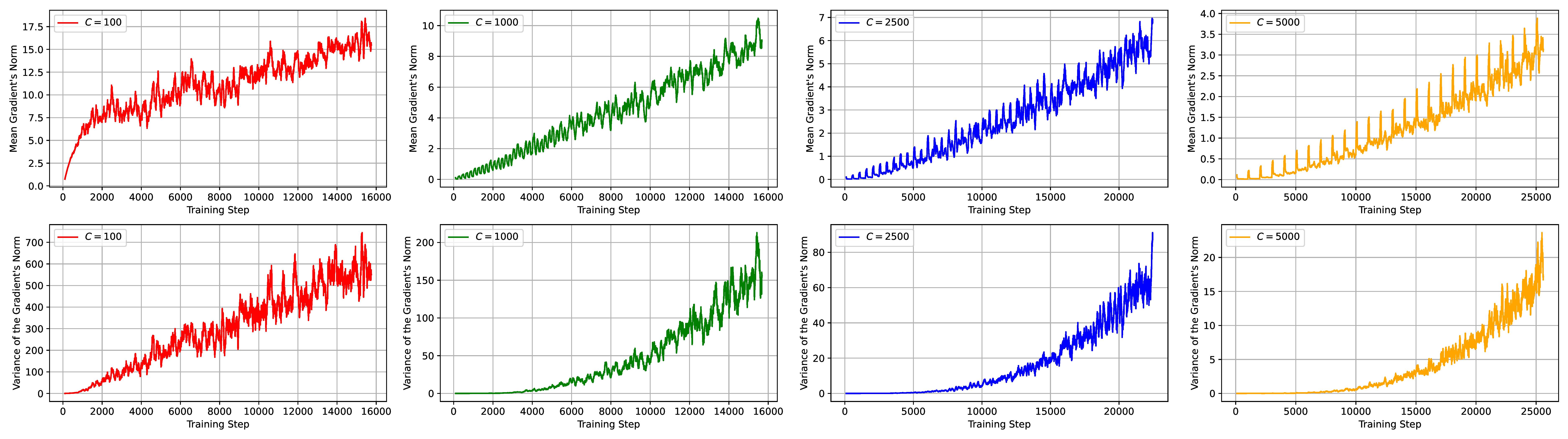}
        \caption{}
        \label{fig:gradients_behavior_acrobot}
    \end{subfigure}
    \caption{Gradients' norm (above) and variance (below) for data re-uploading models with increasing values of $C$ in the Cartpole-v0 (see Subfigure \ref{fig:gradients_behavior_cartpole}) and Acrobot-v1 (see Subfigure \ref{fig:gradients_behavior_acrobot}) environments. }
    \label{fig:gradients_target_network}
\end{figure}

As $C$ increases to substantial values, the models' speed of convergence and performance start to decrease, as expected. However, interestingly, really low values of $C$ (even a value of $C=1$, which corresponds to a model that does not use a target network), achieve a level of performance equal to or even better than moderate values of $C$. In fact, after a search of hyperparameters, \cite{skolik2022quantum} found that $C=1$ leads to the best performance. Turning our attention to the loss functions, we see that, as $C$ increases, the loss function becomes more stable and the maximum values it achieves start decreasing.  Moreover, the dynamics of the moving targets become more evident, as one can see by the loss function for $C=2500$ in the Cartpole-v0 environment, where the peaks correspond to the step the targets change, and the valleys in between them to the steps for which they remain stationary.

Interestingly, both the gradients' norm and variance exhibit a similar behavior to the loss functions, see Figure \ref{fig:gradients_target_network}. This analysis is of utmost importance for VQC-based Deep Q-Learning. In the previous sections, we observed that well-performing models exhibit substantial gradient magnitudes and variances throughout training. This section has empirically demonstrated the influential role of moving targets and the target network update frequency $C$ on the behavior of both the loss function and, consequently, the gradients. In particular, we saw that increasing $C$ stabilizes the loss function, which contributes to more controlled magnitudes of gradients and their variance. However, it is important to highlight that even with relatively low $C$ values — where the gradients' magnitudes and variances were pronounced — the models showcased impressive performance. They were capable of achieving maximum returns in as few episodes as other more "stable" models with higher values of $C$.

This raises an interesting question. It is well known that hardware-efficient VQCs suffer from various trainability issues, from optimization landscapes swamped with bad local minima \cite{anschuetz2022quantum} to the Barren Plateau Phenomenon \cite{mcclean2018barren}, such that the variance of the gradients decays exponentially with system size. However, it appears that Deep Q-Learning is an inherently unstable algorithm due to the moving targets. Intriguingly, certain hyperparameter configurations seem to allow Deep Q-Learning models to effectively learn good policies while sustaining considerable gradient magnitudes and variances. This hints at a potential advantage: might this inherent instability help counteract the aforementioned trainability issues? We have seen that the gradient's variance of the quantum models increases throughout training. Thus, even if these same models suffer from the Barren Plateau Phenomenon, it may be possible to train them in the context of Deep Q-Learning due to the large gradients seen throughout training. To verify these claims, it becomes imperative to research the gradient's behavior for an increasing number of qubits. The most appropriate architecture for such a study is the UQC, since it allows the encoding of an input vector into an arbitrary number of qubits, leading to a higher flexibility in the choice of system size. Thus, in the next section we introduce this architecture and, then, in section \ref{subsection:gradient_qubits}, we analyze how the gradients behave as the number of qubits increases.

\subsection{The Single and Multi-Qubit UQC}\label{subsection:uqc}

We verified in section \ref{subsection:skolik_performance} that the VQC used in \cite{skolik2022quantum} is capable of solving the CartPole-v0 environment and achieving considerable performance in the Acrobot-v1 environment. However, it would be interesting to see if the UQC \cite{perez2020data}, a circuit that also makes use of data re-uploading and trainable input scaling by default, is capable of solving these environments. An advantage of this architecture is the fact that it allows the encoding of any input vector using an arbitrary number of qubits. This raises some interesting questions: Can a single-qubit UQC solve these environments? By increasing the number of qubits, can we improve the performance of the models?

\begin{figure}[h]
    \centering
    \begin{subfigure}[t]{0.8\textwidth}
        \includegraphics[width=\textwidth]{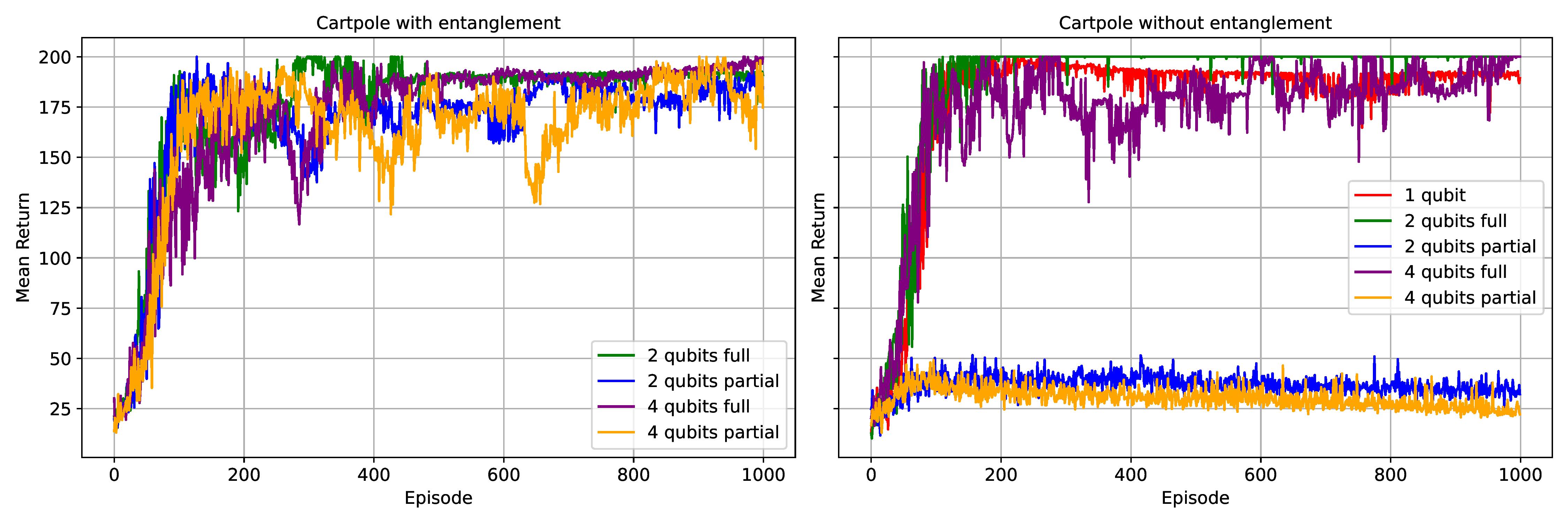}
        \caption{}
        \label{fig:uqc_cartpole}
    \end{subfigure}
    \begin{subfigure}[t]{0.8\textwidth}
        \includegraphics[width=\textwidth]{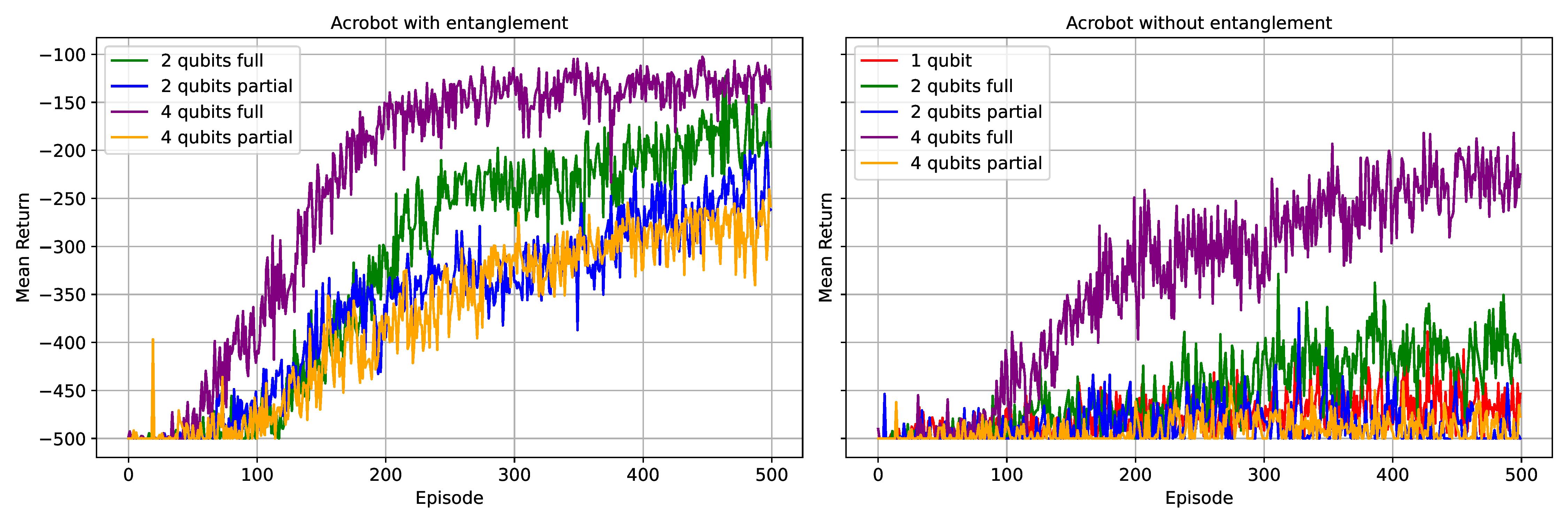}
        \caption{}
        \label{fig:uqc_acrobot}
    \end{subfigure}
    \caption{Performance of UQC models with entanglement (on the left) and without entanglement (on the right) for the CartPole-v0 (see Subfigure \ref{fig:uqc_cartpole}) and the Acrobot-v1 (see Subfigure \ref{fig:uqc_acrobot}) environments. The returns are averaged over $10$ agents. The full set of hyperparameters can be seen in Table \ref{table:uqc}.}
    \label{fig:uqc_entanglement}
\end{figure}

To start, a data encoding technique has to be defined. We experiment with two different types of encoding:
\begin{itemize}
    \item \textbf{Full Encoding}: Using full encoding, the whole input vector is encoded into all the qubits. Consequently, the number of parameters grows linearly with the number of qubits.
    \item \textbf{Partial Encoding}: In this data encoding technique, we divide the number of the input vector features by the number of qubits used and encode a different subvector in each qubit.
\end{itemize}

The reason for using these two types of encoding is two-fold. On the one hand, these two techniques allow us to study the impact of introducing entanglement on the performance of the models in the CartPole-v0 environment. On the other hand, the Full Encoding technique allows us to encode a given input vector in an arbitrary number of qubits that may even be greater than the number of features of the input vector. Thus, we may study how the performance and trainability behave as the number of qubits increases, see Section \ref{subsection:gradient_qubits}. Figures \ref{fig:uqc_cartpole} and \ref{fig:uqc_acrobot} show how each of these data encoding methods perform with and without entanglement for $2$ and $4$ qubits in the Cartpole-v0 and Acrobot-v1 environments, see Subfigures \ref{fig:uqc_cartpole} and \ref{fig:uqc_acrobot}, respectively.

Models without entanglement, that is, models that can be efficiently simulated using classical computers, are able of achieving considerable performance. Moreover, while models with full encoding are capable of learning (sub-)optimal policies, models that use partial encoding require entanglement, such that the features of the input state become correlated, giving the VQC the full information necessary to better approximate the optimal Q-values of the two environments. Finally, even a single-qubit UQC is capable of solving the Cartpole-v0 environment, but the Acrobot-v1 environment needs a VQC with higher width to learn a decent policy.

\subsection{Gradient Behavior for Increasing System Sizes}\label{subsection:gradient_qubits}

One of the advantages of the multi-qubit UQC architecture is that one may encode any input vector into an arbitrary number of qubits. In particular, it is possible to increase the number of qubits even further than the number of features of the input vector. 

\begin{figure}[h]
    \centering
    \includegraphics[width = 0.7\textwidth]{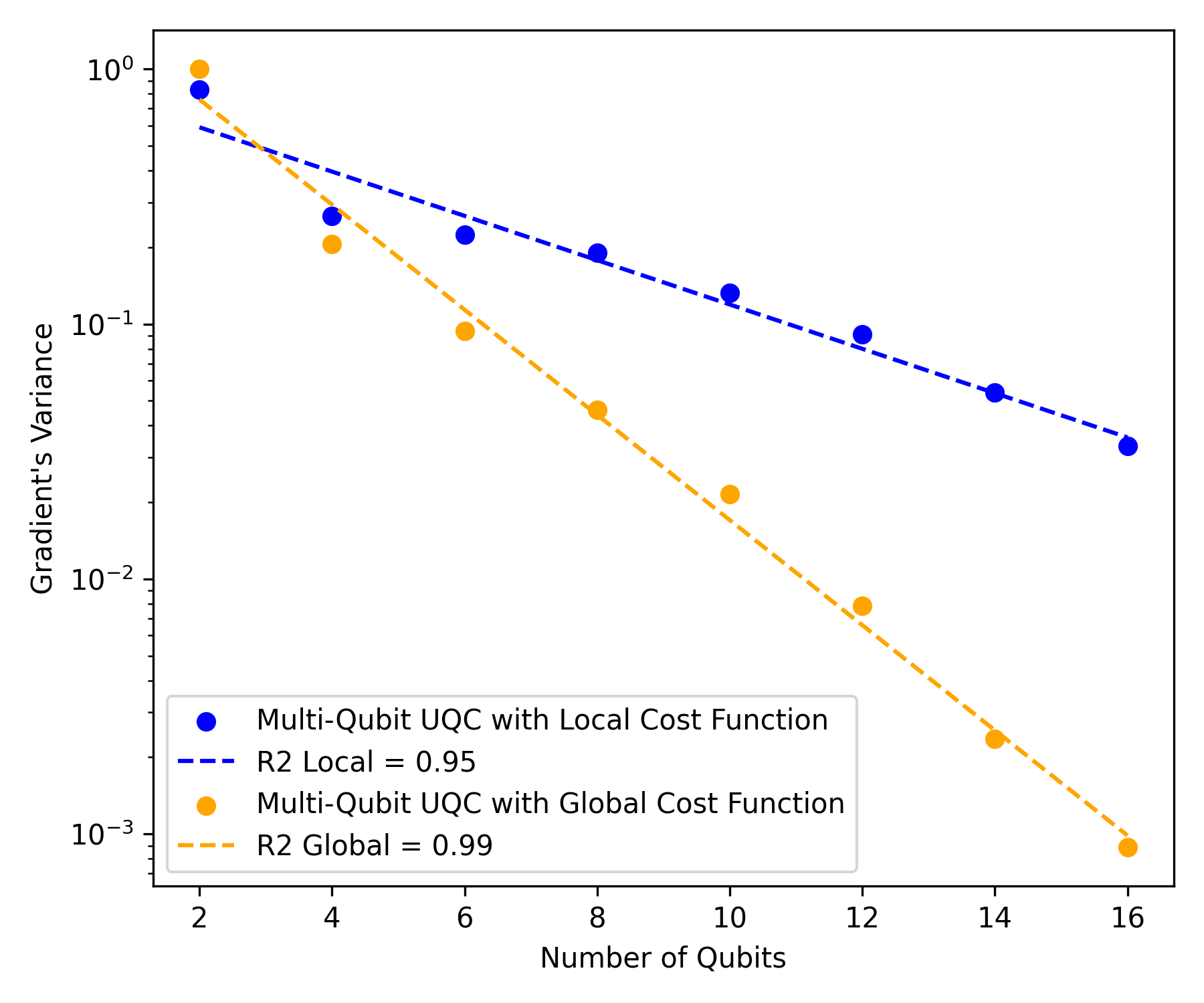}
    \caption{Variance of the Gradient's Norm across $1000$ uniformly sampled parameters spanning the entire Hilbert Space. Two cost functions are tested, a global cost function and a local cost function. For the full set of hyperparameters, see \ref{table: variance_first_step}.}
    \label{fig:variance_first_step}
\end{figure}

We first start by analysing the gradient's variance of the multi-qubit UQC as the number of qubits increases when a uniform initialization that spans the whole Hilbert Space is used. Thus, we only consider the first training step of CartPole-v0 and plot the variance of the norm of the gradient of $1000$ agents for every even number of qubits between $2$ and $16$, see Figure \ref{fig:variance_first_step}.

When a global cost function is used, the variance decays exponentially with the number of qubits. However, for a local cost function (the one used throughout this work for the QRL agents), it seems to decay in a regime on the border of polynomial to exponential (as indicated by the R2 error on the fit), a result consistent with \cite{cerezo2021cost}.

\begin{figure}[h]
    \centering
    \begin{subfigure}[t]{\textwidth}
    \centering
        \includegraphics[width=0.9\textwidth]{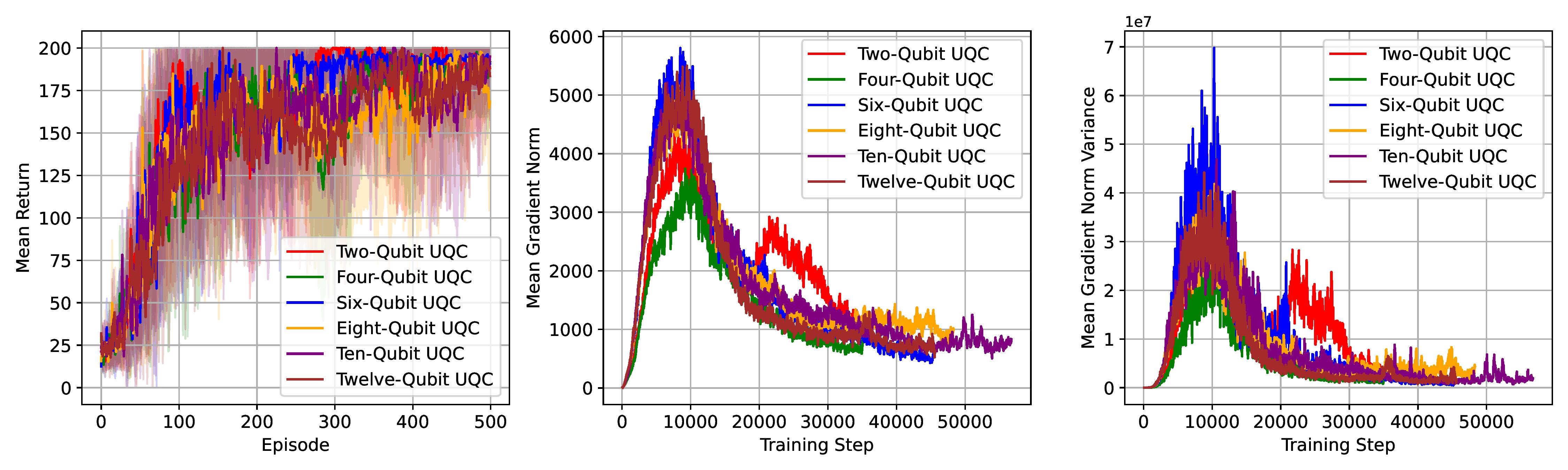}
        \caption{}
        \label{fig:number_of_qubits_cartpole}
    \end{subfigure}
    \begin{subfigure}[t]{\textwidth}
    \centering
        \includegraphics[width=0.9\textwidth]{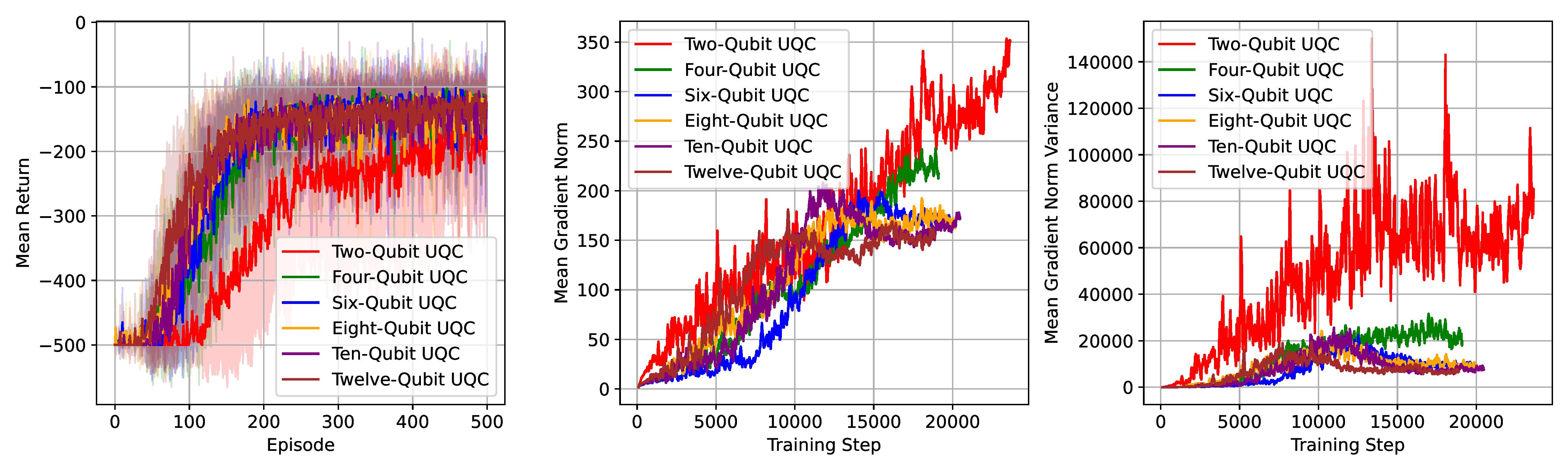}
        \caption{}
        \label{fig:number_of_qubits_acrobot}
    \end{subfigure}
    \caption{Performance (first graph) and Trainability Analysis UQC models with increasing numbers of qubits in the Cartpole-v0 (see Subfigure \ref{fig:number_of_qubits_cartpole}) and Acrobot-v1 (see Subfigure \ref{fig:number_of_qubits_acrobot}) environments. The full set of hyperparameters can be seen in Table \ref{table:gradients_behavior}.}
    \label{fig:number_of_qubits}
\end{figure}

However, we are particularly interested to see how the gradients behave throughout training when using fine-tuned initialization strategies. Thus, we trained the full encoding multi-qubit UQC model with the set of even numbers of qubits from $2$ to $12$ in the CartPole-v0 and Acrobot-v1 environments, see Figures \ref{fig:number_of_qubits_cartpole} and \ref{fig:number_of_qubits_acrobot}, respectively.

Both the magnitude of the gradient and its variance behave similarly during training for all models, while also achieving a similar range of values. However, it is important to note that there seems to be a decrease in the variance of the gradients for the Acrobot-v1 environment as the number of qubits increases, but this seems to be most noticeable after $10000$ training steps, after which most of the training should be over. These results are particularly interesting because they confirm the suspicions that arose throughout this work. Throughout training, all the models exhibit large gradients even as the number of qubits increases. This seems to be similar to warm-starts, since even though these landscapes are characterized by exponentially vanishing gradients, the models are able to navigate the landscape in zones with high gradients, facilitating training. This hints at the possibility of hardware-efficient VQCs being especially suitable to be used as function approximators in Deep Q-learning.

\subsection{Gradient Behavior in Supervised Learning}\label{subsection:supervised}

\begin{figure}[H]
    \centering
    \begin{subfigure}[t]{\textwidth}
    \centering
        \includegraphics[width=0.85\textwidth]{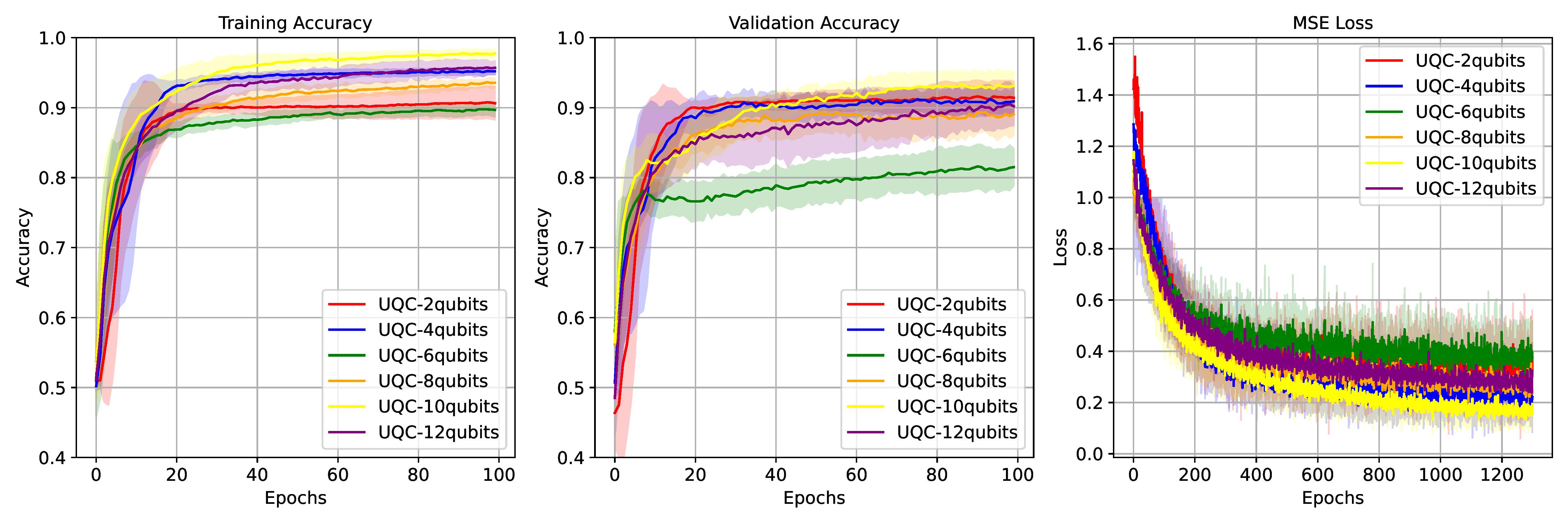}
        \caption{}
        \label{fig:performance_supervised_skolik}
    \end{subfigure}
    \begin{subfigure}[t]{\textwidth}
    \centering
        \includegraphics[width=0.85\textwidth]{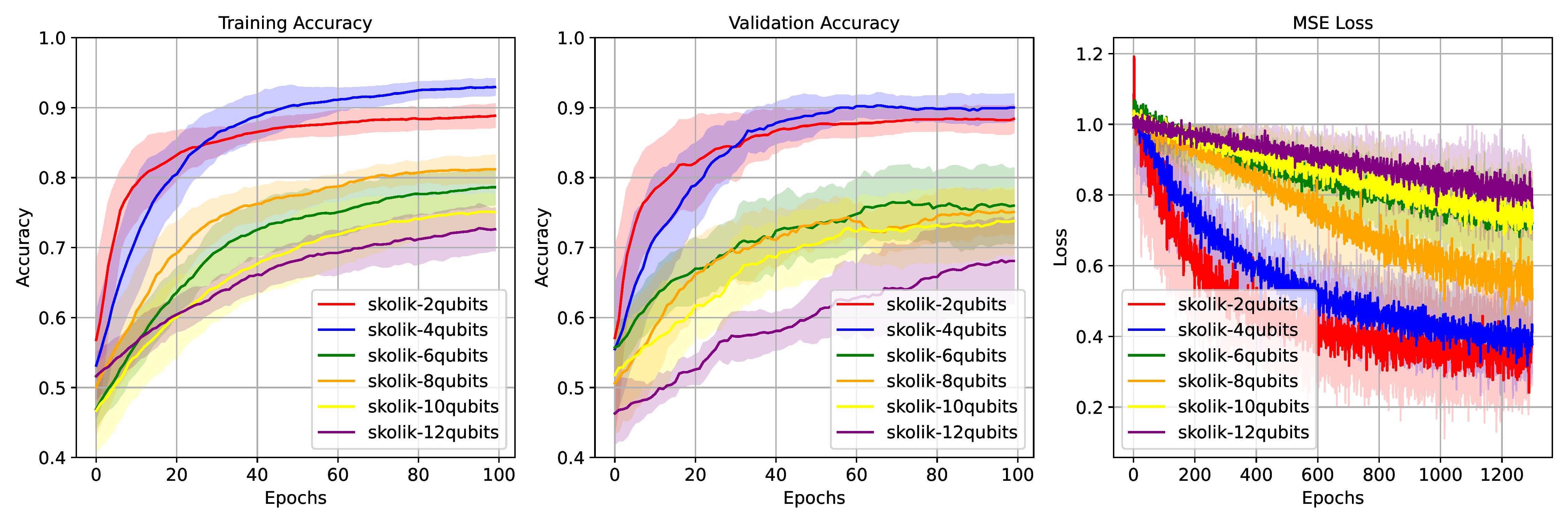}
        \caption{}
        \label{fig:performance_supervised_uqc}
    \end{subfigure}
    \caption{Training and Validation accuracy as well as MSE cost function throughout training for the Skolik Data Re-Uploading (Subfigure \ref{fig:performance_supervised_skolik}) and Full-Encoding Multi-Qubit UQC (see Subfigure \ref{fig:performance_supervised_uqc}) trained on a binary classification problem. The full set of hyperparameters can be seen in \ref{table:supervised}.}
    \label{fig:supervised_performance}
\end{figure}

So far, we have claimed that the high gradients found throughout training by the VQC-based Deep Q-Learning models are due to the instability introduced by the moving targets. To verify whether this is true, we trained both the Skolik Data Re-uploading and the Full Encoding Multi-Qubit UQC on the more common task of binary classification for an increasing number of qubits. Since the Skolik Architecture requires a number of qubits equal to the number of features of the datapoints, we used scipy's \emph{make\_classification} method to generate a dataset of $500$ datapoints for each number of qubits tested. To keep this as close to the original RL scenario as possible, we also used the MSE cost function and the same observables as those used previously. The accuracies achieved on the training and validation datasets as well as the MSE cost function throughout training can be seen in Fig \ref{fig:supervised_performance}

\begin{figure}[H]
    \centering
    \begin{subfigure}[t]{\textwidth}
    \centering
        \includegraphics[width=0.85\textwidth]{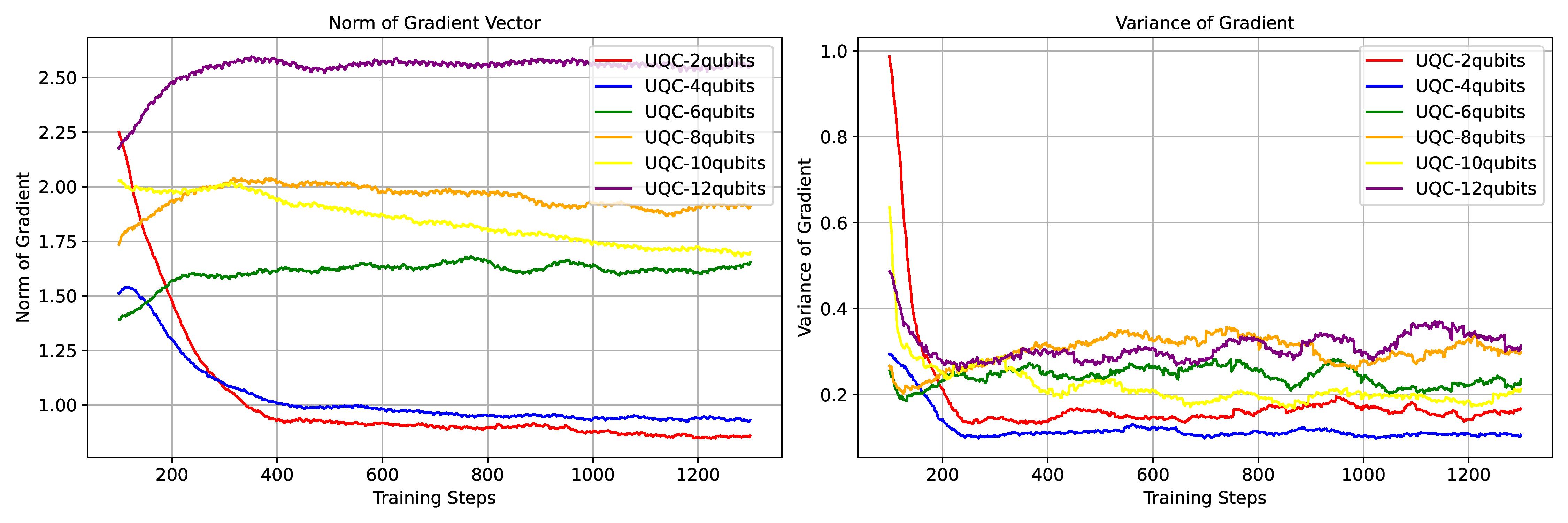}
        \caption{}
        \label{fig:gradients_supervised_skolik}
    \end{subfigure}
    \begin{subfigure}[t]{\textwidth}
    \centering
        \includegraphics[width=0.85\textwidth]{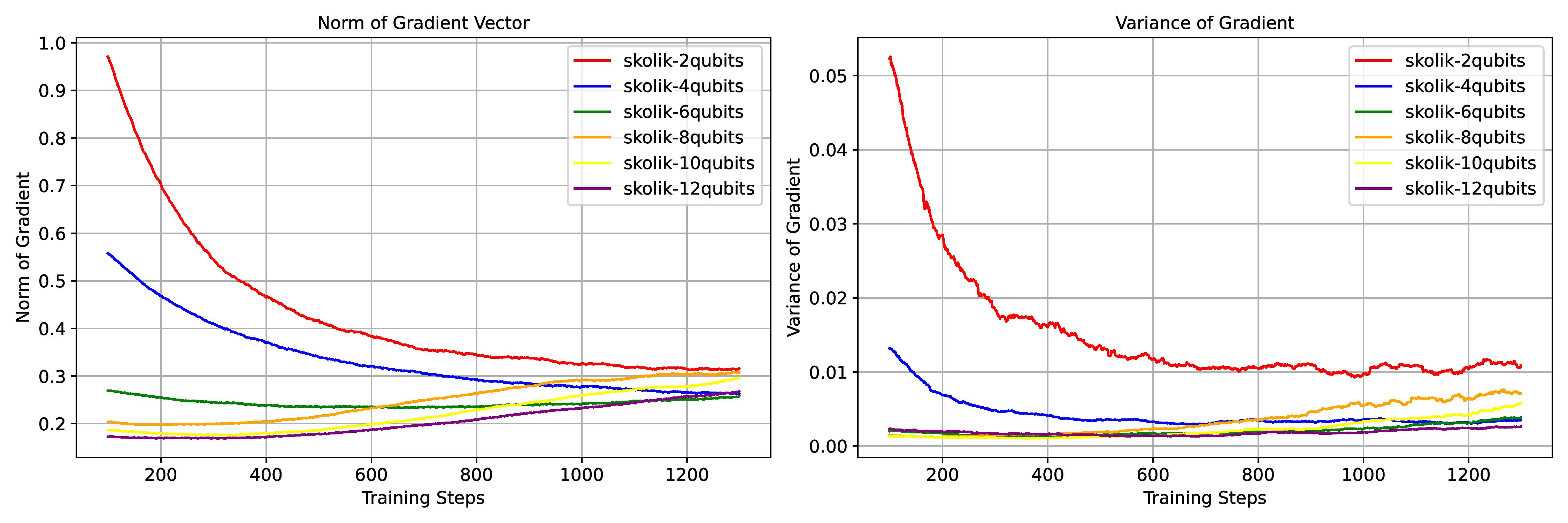}
        \caption{}
        \label{fig:gradients_supervised_uqc}
    \end{subfigure}
    \caption{Norm and Variance of the norm of the gradient vector throughout training for the Skolik Data Re-Uploading (Subfigure \ref{fig:performance_supervised_skolik}) and Full-Encoding Multi-Qubit UQC (see Subfigure \ref{fig:performance_supervised_uqc}) trained on a binary classification problem. The full set of hyperparameters can be seen in \ref{table:supervised}.}
    \label{fig:supervised_gradients}
\end{figure}

Interestingly, the Multi-Qubit UQC model consistently outperforms the Skolik model, achieving higher training and validation accuracies and a lower cost value. Moreover, this performance does not decrease as the number of qubits increases, unlike what happens to the Skolik Model, where for $12$ qubits the validation accuracy does not even reach $70\%$. However, more than the performance of such models in terms of accuracy, we want to know how the gradients behave throughout training for a comparison with what we have seen in the VQC-based Deep Q-Learning agents, see Figure \ref{fig:supervised_gradients}.

The Skolik Models showcase a clear trend in both the norm and the variance of the gradients. As the number of qubits increases, both metrics decrease upon initialization and behave similarly throughout training: they first decrease and then converge. However, for the UQC models, there is not a clear trend in the behavior of the norm of the gradient. Moreover, while the variances across the models with different numbers of qubits behave similarly, there is not a clear decrease in the variance unlike what is seen for the Skolik models.

These results show that the gradients behave differently in the context of supervised learning when compared to Deep Q-learning. This observation confers additional support to the hypothesis that it is the instability caused by the moving targets that allow for high gradients throughout training.

\clearpage

\section{Conclusion}\label{section:conclusion}

The main goal of this work was to study the effects of data re-uploading on the performance and trainability of VQC-based Deep Q-Learning models. It was already empirically shown by \cite{skolik2022quantum} that using data re-uploading increases the performance of these models in the CartPole environment, possibly due to its increased expressivity, which allows the approximation of more intricate functions \cite{schuld2021effect}. Throughout this work, we implemented and tested different data re-uploading models in the CartPole-v0 and Acrobot-v1 environments. Our results match those by \cite{skolik2022quantum} and \cite{schuld2021effect}, re-affirming the effect of data re-uploading on the performance of the models.

However, a concern would be that the increase in expressivity and circuit depth that arises from the use of data re-uploading could decrease the trainability of these models. To investigate such a concern, we analyzed the norm of the gradients and the variance of this norm. From this analysis, we verified that the gradients achieve substantial values and actually increase when data re-uploading is used versus when it is not used. Furthermore, we empirically showed that this increase is due to an instability that is inherent to Deep Q-Learning, which is the fact that the targets are non-stationary. Alongside the Mean Squared Error loss function, this leads to a very unstable algorithm with substantial gradients. Moreover, we also verified that increasing the number of qubits of the multi-qubit UQC in the CartPole-v0 and Acrobot-v1 environments does not lead to a decrease in the gradients' magnitude or variance thorughout training.

\subsection{Future Work}

For future work, it would be interesting to develop a different methodology to analyze the trainability of VQC-based models. The magnitude and variance of the gradients already give some insights into the trainability of a model, but there are other methodologies that could also be used. For instance, it may be possible to study the eigenvalues of the Hessian Matrix.

Another interesting possibility would be to analyze these models from a Fourier Analysis perspective. \cite{schuld2021effect} show that VQCs may be seen as Partial Fourier Series in the data and that using data re-uploading increases the frequencies the VQC "has access to". Thus, it may be possible to analyze the optimal Q-functions of certain environments and the Fourier Series that the VQCs are approximating. This might reveal some insights into what is happening behind the scenes.

Finally, it would be interesting to build upon this work and analyze the trainability of VQC-based Deep Q-Learning models in different, more complex environments, with different loss functions and sets of hyperparameters. It might even be possible to theoretically derive bounds for the gradients of these models, which could reinforce the empirical analysis.

\section*{Experiments Replication}
All data necessary for replicating the results of this paper is available \href{https://github.com/RodrigoCoelho7/VQC_Qlearning}{here}.

\section*{Acknowledgments}
This work is financed by National Funds through the Portuguese funding agency, FCT - Fundação para a Ciência e a Tecnologia, within project LA/P/0063/2020 (\href{https://doi.org/10.54499/LA/P/0063/2020}{https://doi.org/10.54499/LA/P/0063/2020}). RC also thanks the support of the Foundation for Science and Technology (FCT, Portugal) under grant 10053/BII-E\_B4/2023. AS also thanks the support of the Foundation for Science and Technology (FCT, Portugal) within grant UI/BD/152698/2022 and project IBEX, with reference PTDC/CC1-COM/4280/2021.

\newpage
\bibliography{sn-bibliography}
\newpage

\begin{appendices}

\section{Hyperparameters}\label{secA1}

Some of the hyperparameters for the VQC-based Deep Q-Learning models are explained in the following table:

\begin{table}[h]
    \caption{An explanation of VQC-Based Deep Q-Learning's hyperparameters}
    \label{tab:my_label}
    \begin{tabular}{@{}ll@{}}
    \toprule
     \textbf{Hyperparameter} & \textbf{Explanation}\\
     \midrule
     qubits (n) & The quantum circuit's number of qubits\\
     layers & The quantum circuit's number of layers\\
     $\gamma$ & The return's discount factor\\
     trainable input scaling & whether trainable input scaling is used or not\\
     trainable output scaling & whether trainable output scaling is used or not\\
     learning rate of parameters $\theta$ & the learning rate of parameters $\theta$\\
     learning rate of input scaling parameters & the learning rate of input scaling parameters\\
     Learning rate of output scaling parameters & the learning rate of output scaling parameters\\
     batch size & the batch size\\
     decaying schedule of $\epsilon$-greedy policy & defines the decaying schedule of the $\epsilon$-greedy policy (e.g exponential)\\
     $\epsilon_{init}$ & the initial value of $\epsilon$\\
     $\epsilon_{dec}$ & the decay rate of $epsilon$ per episode\\
     $\epsilon_{min}$ & the minimum value of $\epsilon$\\
     update model & the update model's frequency\\
     update target model & the target model's update frequency\\
     data re-uploading & whether data re-uploading is used or not\\
     input scaling initialization & how the input scaling parameters were initialized (Skolik Model)\\
     output scaling initialization & how the output scaling parameters were initialized (Skolik Model)\\
     rotational parameters initialization & how the rotational parameters were initialized ($\theta$ in Skolik, $\varphi$ in UQC)\\
     $\Vec{w}$ initialization & how the $\Vec{w}$ parameters were initialized (UQC)\\
     $\Vec{b}$ initialization & how the $\Vec{b}$ parameters were initialized (UQC)\\
     observables & observables measured\\
     \botrule
    \end{tabular}
\end{table}

The set of hyperparameters used for the VQCs in Figures \ref{fig:skolik_cartpole}, \ref{fig:skolik_acrobot}, \ref{fig:skolik_gradients_cartpole} and \ref{fig:skolik_gradients_acrobot}, can be seen in table \ref{table: skolik_hyper}.

\begin{table}[h]
\caption{Models' hyperpameters of Subsection \ref{subsection:skolik_performance}.}
\label{table: skolik_hyper}
\begin{tabular}{@{}lll@{}}
\toprule
& \textbf{CartPole-v0} & \textbf{Acrobot-v1} \\
\midrule
qubits (n) & $4$ & $4$ \\
layers & $5$ & $5$\\
$\gamma$ & $0.99$ & $0.99$ \\
trainable input scaling & yes, no & yes, no\\
trainable output scaling & yes, no & yes, no\\
learning rate of parameters $\theta$ & $0.001$ & $0.001$\\
learning rate of input scaling parameters & $0.001$ & $0.001$\\
learning rate of output scaling parameters & $0.1$ & $0.1$\\
batch size & $16$ & $32$\\
decaying schedule of $\epsilon$-greedy policy & Exponential & Exponential\\
$\epsilon_{init}$ & $1$ & $1$\\
$\epsilon_{dec}$ & $0.99$ & $0.99$\\
$\epsilon_{min}$ & $0.01$ & $0.01$\\
update model & $1$ & $5$\\
update target model & $1$ & $250$\\
size of replay buffer & $10000$ & $50000$\\
data re-uploading & yes, no & yes, no\\
input scaling initialization & Initialized as $1$s & Initialized as $1$s\\
output scaling initialization & Initialized as $1$s & Initialized as $1$s\\
rotational parameters initialization & Uniformly sampled between $0$ and $\pi$ & Uniformly sampled between $0$ and $\pi$\\
$\Vec{w}$ initialization & -------------- & --------------\\
$\Vec{b}$ initialization & -------------- & --------------\\
observables & $(Z_0Z_1,Z_2Z_3)$ & $(Z_0, Z1Z_2,Z_3)$\\
\botrule
\end{tabular}
\end{table}

The set of hyperparameters used for the VQCs in Figures \ref{fig:target_loss_cartpole} and \ref{fig:target_loss_acrobot}, can be seen in table \ref{table:target_loss}.
\begin{table}[h]
\caption{Models' hyperpameters of Subsection \ref{subsection:tradeoff}.}
\label{table:target_loss}
\begin{tabular}{@{}lll@{}}
\toprule
& \textbf{CartPole-v0} & \textbf{Acrobot-v1} \\
\midrule
qubits (n) & $4$ & $4$ \\
layers & $5$ & $5$\\
$\gamma$ & $0.99$ & $0.99$ \\
trainable input scaling & yes & yes\\
trainable output scaling & yes & yes\\
learning rate of parameters $\theta$ & $0.001$ & $0.001$\\
learning rate of input scaling parameters & $0.001$ & $0.001$\\
learning rate of output scaling parameters & $0.1$ & $0.1$\\
batch size & $16$ & $32$\\
decaying schedule of $\epsilon$-greedy policy & Exponential & Exponential\\
$\epsilon_{init}$ & $1$ & $1$\\
$\epsilon_{dec}$ & $0.99$ & $0.99$\\
$\epsilon_{min}$ & $0.01$ & $0.01$\\
update model & $1$ & $5$\\
update target model & $1,500,1000,2500$ & $100,1000,2500,5000$\\
size of replay buffer & $10000$ & $50000$\\
data re-uploading & yes & yes\\
input scaling initialization & Initialized as $1$s & Initialized as $1$s\\
output scaling initialization & Initialized as $1$s & Initialized as $1$s\\
rotational parameters initialization & Uniformly sampled between $0$ and $\pi$ & Uniformly sampled between $0$ and $\pi$\\
$\Vec{w}$ initialization & -------------- & --------------\\
$\Vec{b}$ initialization & -------------- & --------------\\
observables & $(Z_0Z_1,Z_2Z_3)$ & $(Z_0,Z_1Z_2,Z_3)$\\
\botrule
\end{tabular}
\end{table}

The set of hyperparameters used for the VQCs in Figures \ref{fig:uqc_cartpole} and \ref{fig:uqc_acrobot}, can be seen in table \ref{table:uqc}.
\begin{table}[h]
\caption{Models' hyperpameters of Subsection \ref{subsection:uqc}}
\label{table:uqc}
\begin{tabular}{@{}lll@{}}
\toprule
& \textbf{CartPole-v0} & \textbf{Acrobot-v1} \\
\midrule
qubits (n) & $1,2,4$ & $1,2,4$ \\
layers & $5$ & $5$\\
$\gamma$ & $0.99$ & $0.99$ \\
trainable input scaling & yes & yes\\
trainable output scaling & yes & yes\\
learning rate of parameters $\theta$ & $0.001$ & $0.001$\\
learning rate of input scaling parameters & $0.001$ & $0.001$\\
learning rate of output scaling parameters & $0.1$ & $0.1$\\
batch size & $16$ & $32$\\
decaying schedule of $\epsilon$-greedy policy & Exponential & Exponential\\
$\epsilon_{init}$ & $1$ & $1$\\
$\epsilon_{dec}$ & $0.99$ & $0.99$\\
$\epsilon_{min}$ & $0.01$ & $0.01$\\
update model & $1$ & $5$\\
update target model & $1,500,1000,2500$ & $100,1000,2500,5000$\\
size of replay buffer & $10000$ & $50000$\\
data re-uploading & yes & yes\\
input scaling initialization & -------------- & --------------\\
output scaling initialization & Initialized as $1$s & Initialized as $1$s\\
rotational parameters initialization & Uniformly sampled between $0$ and $\pi$ & Uniformly sampled between $0$ and $\pi$\\
$\Vec{w}$ initialization & Gaussian Distribution (mean=0,std=0.01) & Gaussian Distribution (mean=0,std=0.01)\\
$\Vec{b}$ initialization & Initialized as $0$s & Initialized as $0$s\\
observables & $(Z_0Z_1,Z_2Z_3)$ & $(Z_0,Z_1Z_2,Z_3)$\\
\botrule
\end{tabular}
\end{table}

The set of hyperparameters used for the VQCs in Figures \ref{fig:gradients_behavior_cartpole} and \ref{fig:gradients_behavior_acrobot}, can be seen in table \ref{table:gradients_behavior}.
\begin{table}[h]
\caption{Models' hyperpameters of Subsection \ref{subsection:gradient_qubits}.}
\label{table:gradients_behavior}
\begin{tabular}{@{}lll@{}}
\toprule
& \textbf{CartPole-v0} & \textbf{Acrobot-v1} \\
\midrule
qubits (n) & $2,4,6,8,10,12$ & $2,4,6,8,10,12$ \\
layers & $5$ & $5$\\
$\gamma$ & $0.99$ & $0.99$ \\
trainable input scaling & yes & yes\\
trainable output scaling & yes & yes\\
learning rate of parameters $\theta$ & $0.001$ & $0.001$\\
learning rate of input scaling parameters & $0.001$ & $0.001$\\
learning rate of output scaling parameters & $0.1$ & $0.1$\\
batch size & $16$ & $32$\\
decaying schedule of $\epsilon$-greedy policy & Exponential & Exponential\\
$\epsilon_{init}$ & $1$ & $1$\\
$\epsilon_{dec}$ & $0.99$ & $0.99$\\
$\epsilon_{min}$ & $0.01$ & $0.01$\\
update model & $1$ & $5$\\
update target model & $1,500,1000,2500$ & $100,1000,2500,5000$\\
size of replay buffer & $10000$ & $50000$\\
data re-uploading & yes & yes\\
input scaling initialization & -------------- & --------------\\
output scaling initialization & Initialized as $1$s & Initialized as $1$s\\
rotational parameters initialization & Uniformly sampled between $0$ and $\pi$ & Uniformly sampled between $0$ and $\pi$\\
$\Vec{w}$ initialization & Gaussian Distribution (mean=0,std=0.01) & Gaussian Distribution (mean=0,std=0.01)\\
$\Vec{b}$ initialization & Initialized as $0$s & Initialized as $0$s\\
observables & $(Z_0...Z_{n/2-1},Z_{n/2}...Z_n)$ & $(Z_0,Z_1...Z_{n-1},Z_n)$\\
\botrule
\end{tabular}
\end{table}

\begin{table}[h]
\caption{Models' hyperpameters of Subsection \ref{subsection:gradient_qubits} - Plot \ref{fig:variance_first_step}.}
\label{table: variance_first_step}
\begin{tabular}{@{}lll@{}}
\toprule
& \textbf{UQC Local} & \textbf{UQC Global} \\
\midrule
qubits (n) & $2,4,6,8,10,12,14,16$ & $2,4,6,8,10,12,14,16$ \\
layers & $5$ & $5$\\
$\gamma$ & $0.99$ & $0.99$ \\
trainable input scaling & yes, no & yes, no\\
trainable output scaling & yes, no & yes, no\\
learning rate of parameters $\theta$ & $0.001$ & $0.001$\\
learning rate of input scaling parameters & $0.001$ & $0.001$\\
learning rate of output scaling parameters & $0.1$ & $0.1$\\
batch size & $16$ & $16$\\
decaying schedule of $\epsilon$-greedy policy & Exponential & Exponential\\
$\epsilon_{init}$ & $1$ & $1$\\
$\epsilon_{dec}$ & $0.99$ & $0.99$\\
$\epsilon_{min}$ & $0.01$ & $0.01$\\
update model & $1$ & $1$\\
update target model & $1$ & $1$\\
size of replay buffer & $10000$ & $10000$\\
data re-uploading & yes & yes\\
input scaling initialization & ------------- & -------------\\
output scaling initialization & ------------- & -------------\\
rotational parameters initialization & Uniformly sampled between $0$ and $2\pi$ & Uniformly sampled between $0$ and $2\pi$\\
$\Vec{w}$ initialization & Uniformly between $0$ and $2\pi$ & Uniformly between $0$ and $2\pi$\\
$\Vec{b}$ initialization & Uniformly between $0$ and $2\pi$ & Uniformly between $0$ and $2\pi$\\
observables & $(Z_0...Z_{n/2-1},Z_{n/2}...Z_n)$ & $(Z_0...Z_{n},X_0...Z_n)$\\
\botrule
\end{tabular}
\end{table}

The set of hyperparameters used for the VQCs in Figures \ref{fig:supervised_performance} and \ref{fig:supervised_gradients}, can be seen in table \ref{table:gradients_behavior}.
\begin{table}[h]
\caption{Models' hyperpameters of Subsection \ref{subsection:supervised}.}
\label{table:supervised}
\begin{tabular}{@{}lll@{}}
\toprule
& \textbf{UQC} & \textbf{Skolik} \\
\midrule
qubits (n) & $2,4,6,8,10,12$ & $2,4,6,8,10,12$ \\
layers & $5$ & $5$\\
$\gamma$ & ----------- & ----------- \\
trainable input scaling & yes & yes\\
trainable output scaling & yes & yes\\
learning rate of parameters $\theta$ & $0.001$ & $0.001$\\
learning rate of input scaling parameters & $0.001$ & $0.001$\\
learning rate of output scaling parameters & $0.1$ & $0.1$\\
batch size & $32$ & $32$\\
decaying schedule of $\epsilon$-greedy policy & ----------- & -----------\\
$\epsilon_{init}$ & ----------- & ----------- \\
$\epsilon_{dec}$ & ----------- & ----------- \\
$\epsilon_{min}$ & ----------- & ----------- \\
update model & ----------- & -----------\\
update target model & ----------- & -----------\\
size of replay buffer & ----------- & -----------\\
data re-uploading & yes & yes\\
input scaling initialization & -------------- & Initialized as $1$s\\
output scaling initialization & Initialized as $1$s & Initialized as $1$s\\
rotational parameters initialization & Uniformly sampled between $0$ and $\pi$ & Uniform Sampling between $0$ and $\pi$\\
$\Vec{w}$ initialization & Gaussian Distribution (mean=0,std=0.01) & -----------\\
$\Vec{b}$ initialization & Initialized as $0$s & -----------\\
observables & $(Z_0...Z_{n/2-1},Z_{n/2}...Z_n)$ & $(Z_0...Z_{n/2-1},Z_{n/2}...Z_n)$\\
\botrule
\end{tabular}
\end{table}

\end{appendices}



\end{document}